\documentclass[conference,twocolumn]{IEEEtran}
\ifCLASSINFOpdf
\else
\fi

\usepackage{amsmath, amssymb, amsthm,algorithm,algpseudocode}

\usepackage{tikz}
\usetikzlibrary{shapes,arrows,fit}
\usetikzlibrary{plotmarks}
\usetikzlibrary{positioning}
\usetikzlibrary{shapes.geometric, intersections}
\usetikzlibrary{decorations.markings}

\usepackage{pgfplots}
 \pgfplotsset{compat=newest} 
 \pgfplotsset{plot coordinates/math parser=false}
 
\usepackage[none]{hyphenat}
\usepackage[normalem]{ulem} 
\usepackage{cite}

\usepackage{todonotes}
\usepackage{bbold}

\usepackage{enumitem}

\newcommand{\mc}[1]{\mathcal{#1}}

\newcommand{\truth}{\mbox{\textbb{1}}}

\newcommand{\G}{\text{G}}
\newcommand{\B}{\text{B}}
\newcommand{\GG}{\text{GG}}
\newcommand{\GB}{\text{GB}}
\newcommand{\BG}{\text{BG}}
\newcommand{\BB}{\text{BB}}

\newcommand{\ve}[1]{\uline{\smash #1}}

\newcommand{\Rxone}{\textnormal{Rx$_1$}}
\newcommand{\Rxtwo}{\textnormal{Rx$_2$}}
\newcommand{\Rxj}{\textnormal{Rx$_j$}}

\newcommand{\epsone}[1]{\ensuremath{\epsilon_1(#1)}}			
\newcommand{\epstwo}[1]{\ensuremath{\epsilon_2(#1)}} 			
\newcommand{\epsj}[1]{\ensuremath{\epsilon_j(#1)}}			
\newcommand{\epsonetwo}[1]{\ensuremath{\epsilon_{12}(#1)}}		
\newcommand{\epsonenottwo}[1]{\ensuremath{\epsilon_{1\bar{2}}(#1)}}	
\newcommand{\epsnotonetwo}[1]{\ensuremath{\epsilon_{\bar{1}2}(#1)}}	

\newcommand{\flowdiv}[3]{\ensuremath{d_{#1}^{#2}#3}}		

\newcommand{\flowdivRV}[2]{\ensuremath{D_{#1}^{#2}}}		

\DeclareMathOperator*{\argmax}{argmax}
\DeclareMathOperator*{\argmin}{argmin}

\allowdisplaybreaks[4] 

\tikzstyle{chan} = [draw, rectangle, rounded corners,minimum height=2em, minimum width=4em]
\tikzstyle{block} = [draw, rectangle, minimum height=4em, minimum width=2em]
\tikzstyle{block2} = [draw, rectangle, minimum height=1.5em, minimum width=2em]
\tikzset{->-/.style={decoration={markings,mark=at position #1 with {\arrow{>}}},postaction={decorate}}}

\theoremstyle{definition}

 \newtheorem{theorem}{Theorem}
 \newtheorem{prop}{Proposition}
 \newtheorem{remark}{Remark}
 \newtheorem{lemma}[theorem]{Lemma}
 
 \newtheorem{cor}[theorem]{Corollary}
 \newtheorem{thm}{Theorem}

\begin{document}

\title{Capacity Regions of Two-User Broadcast Erasure Channels with Feedback and Hidden Memory}

\author{\IEEEauthorblockN{Michael Heindlmaier, Shirin Saeedi Bidokhti}
\IEEEauthorblockA{Institute for Communications Engineering, Technische Universit\"at M\"unchen,
Munich, Germany\\
Email: michael.heindlmaier@tum.de, shirin.saeedi@tum.de}}

\maketitle

\begin{abstract}
The two-receiver broadcast packet erasure channel with feedback and memory is studied. Memory is modeled using a finite-state Markov chain representing a channel state. The channel state is unknown at the transmitter, but observations of this hidden Markov chain are available at the transmitter through feedback. Matching outer and inner bounds are derived and the capacity region is determined. The capacity region does not have a single-letter characterization and is, in this sense, uncomputable. Approximations of the capacity region are provided and two optimal coding algorithms are outlined. The first algorithm is a probabilistic coding scheme that bases its decisions on the past $L$ feedback sequences. Its achievable rate-region approaches the capacity region exponentially fast in $L$. The second algorithm is a backpressure-like algorithm that performs optimally in the long run.
\end{abstract}  

\section{Introduction}

In many communication protocols, information transmission is done in a packet by packet manner, and  the receiving devices either correctly receive the transmitted packet or detect an erasure. Packet erasure channels (PECs) model such systems. Broadcast PECs (BPECs), in particular, are interesting models to study the broadcast nature of wireless systems. 

The capacity region of the general broadcast channel (BC) remains unresolved both without and with feedback. It was shown in \cite{gamal1978feedback} that feedback does not increase the capacity  of physically degraded BCs. 
It is known that feedback increases the capacity of general BCs and even partial feedback can help \cite{dueck1980partial, kramer2003capacity,ozarow1984broadcast,4655434}.

Remarkably, capacities of memoryless BPECs are known both with and without feedback for two receivers. The former is a special case of the work in \cite{gallager1974capacity} which characterized the capacity of degraded broadcast channels. The latter was derived in \cite{georgiadis2009broadcast}. Generalizations of the idea presented in \cite{georgiadis2009broadcast} have led to optimal coding schemes for BPECs with three receivers and  near-optimal coding schemes for more receivers in \cite{6522177, gatzianas2012feedback, wang2012capacity}. The schemes are feedback-based coding algorithms that are based on network coding techniques and have also influenced related recent works on continuous output broadcast channels with feedback and fading \cite{song2012network, maddah2012completely}.

This work studies BPECs with feedback and channel memory. 
The problem is motivated by the bursty nature of erasures in practical communication systems, e.g. satellite links \cite{lutz1991land, fontan2001statistical,ibnkahla2004high}. 
When there is no feedback, one can use erasure correcting codes for memoryless channels in combination with interleavers to decorrelate the erasures. Interestingly, feedback enables more sophisticated coding techniques  \cite{heindlmaier2014netcod}. In a related work, \cite{dabora2010capacity} derived an infinite-letter capacity characterization for two special cases of the general BC with feedback, memory, and unidirectional receiver cooperation.

We model the memory of a channel by a finite state machine and a set of state-dependent erasure probabilities. This is a well-studied approach for wireless channels (see \cite[Chapter 4.6]{gallager1968information}, \cite{sadeghi2008finite} and the references therein). 

In a previous work \cite{heindlmaier2014oncapacity}, we studied BPECs with feedback and memory under the assumption that the channel state is causally known at the transmitter. We derived close inner and outer bounds on the capacity region. This problem was also studied in parallel by Kuo and Wang in \cite{Kuo_Wang2014, 6847153}. They proposed a new coding strategy and characterized its corresponding rate region. The outer bound in \cite{heindlmaier2014oncapacity} and the inner bound in \cite{Kuo_Wang2014, 6847153} match and thus characterize the capacity region.
This paper extends the previous results to the case where the channel state is no longer observable at the transmitter (except through the receivers' feedback) and thus evolves according to a hidden Markov model from the transmitter's point of view.

Our contribution in this paper is as follows.
We propose an outer bound on the capacity region that has an $n$-letter characterization in terms of a feasibility problem. This region is not computable because its computation needs an infinitely large $n$, and the  number of parameters in the corresponding feasibility problem grows exponentially in $n$. Nevertheless, under mild conditions, we find a sequence of approximations on this region, $\bar{\mc C}^\text{mem}_\text{fb}(L)$, for every $L\geq1$. The $L^\text{th}$ order approximation of the region approaches the outer bound exponentially fast in $L$. For every $L$, we propose a probabilistic encoding strategy at the transmitter that bases its decisions on the past $L$ feedback symbols and achieves the $L^\text{th}$ order approximation of the outer bound. 
%
Finally, a deterministic algorithm is outlined that bases its decisions on the entire past feedback symbols, is implementable, and optimal in the long run.

This paper is organized as follows. 
We introduce the system model in Sec.~\ref{sec:model}. An outer bound on achievable rates is derived in Sec.~\ref{sec:proof_bound_hidden}. Achievable schemes are discussed in Sec.~\ref{sec:achievability} before we conclude in Sec.~\ref{sec:results}. Proofs and detailed derivations can be found in the Appendix.

\section{Notation and System Model}
\label{sec:model}

\begin{figure}[t]
\centering
\begin{tikzpicture}[node distance=5mm, scale=.5, font=\small]
\input{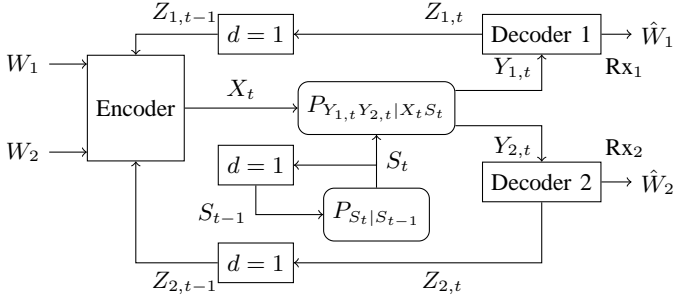}
\end{tikzpicture}
\caption{Block diagram for the BPEC. The box marked with $d=1$ represents a delay of one time unit.}
\label{fig:block}
\vspace*{-5mm}
\end{figure}

\subsection{Notation}
Random variables (RVs) are denoted by capital letters.
A finite sequence (or string) of RVs $X_1,X_1,\ldots,X_n$ is denoted by $X^n$. 
Sequences can have subscripts, e.g. $X_j^n$ is shorthand for $X_{j,1},X_{j,2},\ldots,X_{j,n}$.
Vectors are written with underlined letters, e.g.,  $\ve Z_t=(Z_{1,t}, Z_{2,t})$.
Sets are denoted by calligraphic letters, e.g., $\mc X$.
The indicator function $\truth\{\cdot \}$ takes on the value $1$ if the event inside the brackets is true and $0$ otherwise. 
Conditional probabilities are written equivalently as $\Pr[X=x|Y=y]$, $P_{X|Y}(x|y)$ or sometimes $P(x|y)$ if the involved RVs are clear from the context.
The conditional expectation of a function $f$ of a random variable $X$ given another random variable $Z$ is itself a random variable and is written as $\mathbb E[f(X)|Z]$. Using the law of total expectation $\mathbb E[f(X)|Z] = \mathbb E\big[\mathbb E[f(X)|YZ] \big| Z \big]$. Note that if $X-Y-Z$ forms a Markov chain, we can write $\mathbb E[f(X)|Z] = \mathbb E\big[\mathbb E[f(X)|Y] \big| Z \big]$.

\subsection{System Model}
A transmitter wishes to communicate two independent messages $W_1$ and $W_2$ (of $nR_1$, $nR_2$ packets, respectively) to two receivers $\text{Rx}_1$ and $\text{Rx}_2$ over $n$ channel uses. Communication takes place over a BPEC with memory and feedback as shown in Fig. \ref{fig:block} and described below.

The input to the BPEC at time $t$ is denoted by $X_t \in \mc X$, $t=1,\ldots,n$.
The channel inputs are packets of $\ell$ bits; i.e. $\mc X = \mathbb{F}_{q}$ with $q=2^\ell$, and $\ell \gg 1$.
All rates are measured in packets per slot and entropies and mutual information terms are with respect to logarithms to the base $q$. 

The channel output at \Rxj~at time $t$ is written as $Y_{j,t}\in\mathcal{Y}$, $j\in\{1,2\}$, where $\mathcal{Y}=\mc X \cup \{E\}$.  
Each $Y_{j,t}$ is either $X_t$ (received perfectly) or $E$ (erased).

We define binary RVs $Z_{j,t}$, $j\in\{1,2\}$, $t=1,\ldots,n$, to indicate erasure at \Rxj~at time $t$; i.e. $Z_{j,t}=\truth\{Y_{j,t}=E\}$.
$Y_{j,t}$ can be expressed as a function of $X_{t}$ and $Z_{j,t}$ but $Y_{j,t}$ also determines $Z_{j,t}$. 
We collect $(Z_{1,t},Z_{2,t})$ into the vector $\ve Z_t$, with $\ve Z_t \in \mc Z$, $\mc Z = \{(0,0),(0,1),(1,0),(1,1)\}$.

The memory of the BPEC under study is modelled with a finite state machine with state $S_t$ at time $t$. 
The underlying homogeneous finite state Markov chain is assumed to be irreducible and aperiodic with state space $\mc S$ and steady-state distribution $\pi_s$, $s \in \mc S$. The initial state $S_0$ is distributed according to the steady-state distribution, so the sequence $S^n$ is stationary.
The state-dependent erasure probabilities are specified through
$P_{\ve Z_{t}|S_{t}}$ that does not depend on $t$.
We permit arbitrary correlation between $Z_{1,t}$ and $Z_{2,t}$.
The erasures $\ve Z^n$ are correlated in time in general, hence the channel has memory.
The sequence $\ve Z^n$ is stationary due to stationarity of $S^n$.

After each transmission, an ACK or negative ACK (NACK) feedback is available at the encoder from both receivers. As opposed to our previous work in \cite{heindlmaier2014oncapacity}, we assume no separate feedback of the channel state; i.e., the channel state is \emph{not} known at the transmitter. The channel input at time $t$ may be written as
\begin{align}
X_t = f_t(W_1,W_2,\ve Z^{t-1}).
\end{align}

The probability of erasure in each channel state is described based on the past channel state. Nevertheless, at the transmitter, the channel state is not available and thus the probabilities of erasure events depend on all the past observed feedback messages, $\ve z^{t-1}$, as follows:
\begin{align}
 \epsonetwo{\ve z^{t-1}}&=P_{\ve Z_t|\ve Z^{t-1}}(1,1|\ve z^{t-1})\nonumber\\
 \epsilon_{\bar 12}(\ve z^{t-1})&=P_{\ve Z_t|\ve Z^{t-1}}(0,1|\ve z^{t-1})\nonumber\\
 \epsilon_{1\bar 2}({\ve z^{t-1}})&=P_{\ve Z_t|\ve Z^{t-1}}(1,0|\ve z^{t-1})\nonumber\\
 \epsone{\ve z^{t-1}}&=\epsonetwo{\ve z^{t-1}}+\epsonenottwo{\ve z^{t-1}}\nonumber\\
 \epstwo{\ve z^{t-1}}&=\epsonetwo{\ve z^{t-1}}+\epsnotonetwo{\ve z^{t-1}} 
   \label{eq:def_epsilon_hmm}
\end{align}

Each decoder $\text{Rx}_j$ has to reliably estimate $\hat W_j$ from its received sequence $Y_j^n$.
A rate-pair $(R_1,R_2)$ is said to be achievable if the error probability $\Pr[\hat W_1 \neq W_1, \hat{W}_2\neq W_2]$ can be made to approach zero as $n$ gets large.
The capacity region $\mc C_{\text{ fb}}^\text{mem}$ is the closure of the achievable rate pairs.

 \section{The Outer Bound}
 \label{sec:proof_bound_hidden}

Define $\bar{\mc C}_{n,\text{fb}}^\text{mem}$, for every integer $n$, as the closure of all rate pairs $(R_1,R_2)$ for which there exist variables $x(\ve z^{t-1}), y(\ve z^{t-1})$, $\ve z^{t-1}\in \mc Z^{t-1}$, $t=1,\ldots,n$, such that
\begin{align}
&0\leq x(\hspace{-.025cm}\ve z^{t\!-\!1}\hspace{-.025cm}),\, y(\hspace{-.025cm}\ve z^{t\!-\!1}\hspace{-.025cm}) \!\leq\! 1,\   \forall{t\!=\!1,\ldots,n,\ {\forall\!~\ve z^{t\!-\!1}\!\!\in\!\mc Z^{t\!-\!1}}}\label{eq:posouter_hmm}\\
&R_1\!\leq\!\frac{1}{n} \sum_{t=1}^n \sum_{\ve z^{t\!-\!1}\in\mc Z^{t\!-\!1}}\! \!P_{\ve Z^{t\!-\!1}}(\hspace{-.025cm}\ve z^{t\!-\!1}\hspace{-.025cm}) (1\!-\!\epsone{\hspace{-.025cm}\ve z^{t\!-\!1}\hspace{-.025cm}}\!) x(\hspace{-.025cm}\ve z^{t\!-\!1}\hspace{-.025cm}) \label{eq:R1_constr1outer_hmm}\\
&R_1\!\leq\!\frac{1}{n} \sum_{t=1}^n \sum_{\ve z^{t\!-\!1}\in\mc Z^{t\!-\!1}} \!\!P_{\ve Z^{t\!-\!1}}\!(\hspace{-.025cm}\ve z^{t\!-\!1}\hspace{-.025cm}) (1\!-\!\epsonetwo{\hspace{-.025cm}\ve z^{t\!-\!1}\hspace{-.025cm}}\!) (1\!\!-\!\!y(\hspace{-.025cm}\ve z^{t\!-\!1}\hspace{-.025cm})\!)\label{eq:R1_constr2outer_hmm} \\
&R_2\!\leq\!\frac{1}{n} \sum_{t=1}^n \sum_{\ve z^{t\!-\!1}\in\mc Z^{t\!-\!1}} \!\!P_{\ve Z^{t\!-\!1}}(\hspace{-.025cm}\ve z^{t\!-\!1}\hspace{-.025cm}) (1\!-\!\epstwo{\hspace{-.025cm}\ve z^{t\!-\!1}\hspace{-.025cm}}\!) y(\hspace{-.025cm}\ve z^{t\!-\!1}\hspace{-.025cm}) \label{eq:R2_constr1outer_hmm}\\
&R_2\!\leq\!\frac{1}{n} \sum_{t=1}^n \sum_{\ve z^{t\!-\!1}\in\mc Z^{t\!-\!1}} \!\!P_{\ve Z^{t\!-\!1}}(\hspace{-.025cm}\ve z^{t\!-\!1}\hspace{-.025cm}) (1\!-\!\epsonetwo{\hspace{-.025cm}\ve z^{t\!-\!1}\hspace{-.025cm}}\!) (1\!\!-\!\!x(\hspace{-.025cm}\ve z^{t\!-\!1}\hspace{-.025cm})\!). \label{eq:R2_constr2outer_hmm}
\end{align}

Define $\bar{\mc C}_{\text{ fb}}^\text{mem}$ as the limsup of $\bar{\mc C}_{n,\text{fb}}^\text{mem}$, when $n\to\infty$.
With similar steps as in \cite[Section IV]{heindlmaier2014oncapacity}, we have:
\begin{thm}
\label{outerbound}
Any achievable rate pair $(R_1,R_2)$ is such that $(R_1-\delta,R_2-\delta)\in \bar{\mc C}_{\text{fb}}^\text{mem}$, for $\delta>0$.
\end{thm}
The proof can be found in Appendix~\ref{ap-outerboundproof}. 
\begin{remark}
To compare this result with the outer bound in \cite{heindlmaier2014oncapacity}, where the channel state was strictly causally known at the transmitter, we note that Theorem~\ref{outerbound} is a very general form of \cite{heindlmaier2014oncapacity} with infinite states. Each state represents the state of the system at time $t$, with a feedback $\ve{Z}^{t-1}$ received.
\end{remark}
\begin{remark}
Consider a Gilbert-Elliot model with state space $\mathcal{S}=\{\GG,\GB,\BG,\BB\}$, where $\G$ and $\B$ refer to a good and a bad channel state at each user, respectively. Consider the special case where we always have an erasure in state $\B$ and no erasure in state $\G$. 
Even when there is no separate channel state feedback, this channel model may be considered as a special case of \cite{heindlmaier2014oncapacity}, for the transmitter has full knowledge about the past channel state upon receiving the feedback. It is not hard to show that, in this special case, Theorem~\ref{outerbound} reduces to the outer bound in \cite{heindlmaier2014oncapacity}.
\end{remark}

Theorem~\ref{outerbound} establishes an outer bound on the capacity region. 
The above characterization is, however, uncomputable. We find an approximation of the region next.

\subsection{Approximation of the Outer Bound}
\label{sec:outer_bounds}
The derived outer bound has an infinite-letter characterization because $S_{t-1}$ is not fed back to the transmitter, but an estimate of it is available through the feedback symbols $\ve Z^{t-1}$. From the transmitter's perspective, the predicted erasure probabilities $P_{\ve Z_t|\ve Z^{t-1}}$ depend on all the past $\ve Z^{t-1}$. 
Intuitively, one expects that the effect of past feedback diminishes rapidly. We thus approximate the infinite-letter bound in \eqref{eq:posouter_hmm}~-~\eqref{eq:R2_constr2outer_hmm} with a finite-letter bound where channel erasure events depend only on the past finite $L$ feedback symbols.
The predicted erasure probabilities for time slot $t$ are as follows:
\begin{align}
 P(\ve z_{t}|\ve z^{t-1}) = \sum_{s \in \mc S} P_{\ve Z_{t}|S_{t}}(\ve z_{t}|s) P_{S_{t}|\ve Z^{t-1}}(s|\ve z^{t-1})
\label{eq:prediction_erasure_prob}
\end{align}
$P_{\ve Z_{t}|S_{t}}$ does not depend on $t$, but $P_{S_{t}|\ve Z^{t-1}}$ does. In practice, one observes that the latter distribution is ``close" to $P_{S_{t}|\ve Z^{t-1}_{t-L}}$ which predicts channel states from the past $L$ feedback symbols if $L$ is reasonably large. 
This is made precise by the following theorem which is adapted from \cite[Theorem 2.1]{le2000exponential}.
\begin{thm}[{cf. \cite[Theorem 2.1]{le2000exponential}}]
\label{thm:hmm_forgetting}
 Suppose that all entries of both the state transition matrix $P_{S_t|S_{t-1}}$ and the distribution matrix $P_{\ve Z_t|S_{t}}$ are strictly positive.
For any observed sequence $\ve z^{t-1}\in \mc Z^{t-1}$, we have
\begin{align}
 \sum_{s \in \mc S} \left| P_{S_{t}|\ve Z^{t-1}}(s|\ve z^{t-1}) -  P_{S_{t}|\ve Z^{t-1}_{t-L}}(s|\ve z^{t-1}_{t-L})    \right| \leq {2  (1-\sigma)^L}, \nonumber
\end{align}
where $\sigma\!>\!0$ depends on the smallest entry of matrix $P_{S_t|S_{t-1}}$ and the ratio of the largest and smallest values in matrix $P_{\!\ve Z_t|S_{t}}$.
$\ve Z_{t-L}^{t-1}$ is shorthand for $\ve Z_{t-L}, \ve Z_{t-L+1}, \ldots, \ve Z_{t-1}$.
\end{thm}

\begin{remark}
 The theorem in its original form \cite[Theorem 2.1]{le2000exponential} is more general and only requires the $r$-th order transition matrix $P_{S_t|S_{t-r}}$ to be strictly positive. We use the weaker form here for simplicity.
\end{remark}
The following corollary, derived in detail in Appendix~\ref{sec:derivation_corollary_forgetting}, is the basis of the approximated outer bound:
\begin{cor}
\label{cor:Pz_decay}
Under the conditions of Theorem~\ref{thm:hmm_forgetting}, the total variation between $P_{\ve Z_{t}|\ve Z^{t-1}}$ and $P_{\ve Z_{t}|\ve Z^{t-1}_{t-L}}$ is bounded for any sequence $\ve z^{t-1}\in\mc Z^{t-1}$ by
 \begin{align}
  \sum_{\ve z_{t}} \left| P(\ve z_{t}|\ve z^{t-1}) -  P(\ve z_{t}|\ve z^{t-1}_{t-L})    \right| \leq 2 (1-\sigma)^L. \label{eq:cor_Pz_decay}
\end{align}
\end{cor}
Hence an approximate characterization of the outer bound $\bar{\mc C}_{\text{ fb}}^\text{mem}$ is given by the following feasibility problem.
\begin{align}
0&\leq x(\ve z^L), y(\ve z^L)\leq 1,  \qquad \forall~{\ve z^L\in\mc Z^L }\label{eq:posouter_hmm_wind_outer}\\
R_1&\leq \sum_{\ve z^L\in\mc Z^L} P_{\ve Z^L}(\ve z^L) (1-\epsone{\ve z^L}) x(\ve z^L) + C_L\label{eq:R1_constr1outer_hmm_wind_outer}\\
R_1&\leq \sum_{\ve z^L\in\mc Z^L} P_{\ve Z^L}(\ve z^L) (1-\epsonetwo{\ve z^L}) (1-y(\ve z^L)) + C_L \label{eq:R1_constr2outer_hmm_wind_outer} \\
R_2&\leq \sum_{\ve z^L\in\mc Z^L} P_{\ve Z^L}(\ve z^L) (1-\epstwo{\ve z^L}) y(\ve z^L) + C_L\label{eq:R2_constr1outer_hmm_wind_outer}\\
R_2&\leq \sum_{\ve z^L\in\mc Z^L} P_{\ve Z^L}(\ve z^L) (1-\epsonetwo{\ve z^L}) (1-x(\ve z^L)) + C_L , \label{eq:R2_constr2outer_hmm_wind_outer}
\end{align}
where $- 2 (1-\sigma)^L\leq C_L\leq 2 (1-\sigma)^L $ and $\epsj{\ve z^L}$, $\epsonetwo{\ve z^L}$ are computed via the distribution $P_{\ve Z_{t}|\ve Z^{t-1}_{t-L}}$.
A detailed derivation can be found in Appendix~\ref{sec:proof_bound_hidden_approx}.

By setting $C_L=0$, we obtain what we call the $L^{\text{th}}$ order approximation of the outer bound and we denote it by $\bar{\mc C}_{\text{ fb}}^\text{mem}(L)$. Clearly, $\bar{\mc C}_{\text{ fb}}^\text{mem}(L)$ approaches $\bar{\mc C}_{\text{ fb}}^\text{mem}$ exponentially fast in $L$.
$\bar{\mc C}_{\text{ fb}}^\text{mem}(L)$ is of finite-letter form and thus computable. The number of parameters in the corresponding feasibility problem is exponential in $L$. Since the approximate rate-region approaches the outer bound exponentially fast in $L$, it can give good approximations for reasonable values of $L$. 
 \begin{remark}
  For many examples, the parameter $\sigma$ in \eqref{eq:cor_Pz_decay} turns out to be close to zero. In order to give meaningful bounds, $L$ needs to grow large (in the order of $L=10000$). Nevertheless, we observe in numerical examples that the variational distance decays much faster than ensured by \eqref{eq:cor_Pz_decay}. 
\end{remark}

 \section{Achievability}
 \label{sec:achievability}
 
We develop codes that achieve the outer bound in Sec.~\ref{sec:proof_bound_hidden}.
 The coding strategy is based on the coding techniques discussed in \cite{Kuo_Wang2014}. More precisely, it uses network coding techniques of \cite{georgiadis2009broadcast, heindlmaier2014oncapacity} and a proactive coding strategy that was proposed in \cite{Kuo_Wang2014} for the problem with causal channel state information. We discuss this coding scheme in our own framework to give an alternative viewpoint and as we believe the analysis is simpler than the one provided in \cite{Kuo_Wang2014}.

\subsection{Queue Model}

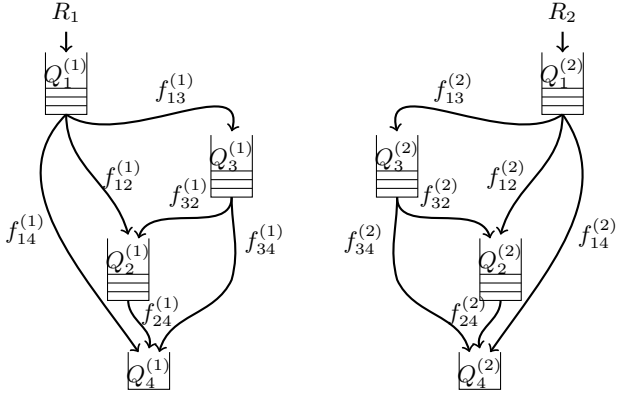
\begin{figure}[t!]
\centering
\begin{tikzpicture}[scale=0.55,font=\small]
\draw (0,0) -- ++(0,-1.5cm) -- ++(1.0cm,0) -- ++(0,1.5cm);
\foreach \i in {1,...,3}
  \draw (0,-1.5cm+\i*6pt) -- +(+1.0cm,0);

\draw (1.5,-4.5) -- ++(0,-1.5cm) -- ++(1.0cm,0) -- ++(0,1.5cm);
\foreach \i in {1,...,3}
  \draw (1.5,-6cm+\i*6pt) -- +(+1.0cm,0);

\draw (4,-2) -- ++(0,-1.5cm) -- ++(1.0cm,0) -- ++(0,1.5cm);
\foreach \i in {1,...,3}
  \draw (4,-3.5cm+\i*6pt) -- +(+1.0cm,0);

\draw (2,-7.25) -- ++(0,-0.9cm) -- ++(1.0cm,0) -- ++(0,0.9cm);

\draw (12,0) -- ++(0,-1.5cm) -- ++(1.0cm,0) -- ++(0,1.5cm);
\foreach \i in {1,...,3}
  \draw (12,-1.5cm+\i*6pt) -- +(+1.0cm,0);

\draw (10.5,-4.5) -- ++(0,-1.5cm) -- ++(1.0cm,0) -- ++(0,1.5cm);
\foreach \i in {1,...,3}
  \draw (10.5,-6cm+\i*6pt) -- +(+1.0cm,0);

\draw (8,-2) -- ++(0,-1.5cm) -- ++(1.0cm,0) -- ++(0,1.5cm);
\foreach \i in {1,...,3}
  \draw (8,-3.5cm+\i*6pt) -- +(+1.0cm,0);

\draw (10,-7.25) -- ++(0,-0.9cm) -- ++(1.0cm,0) -- ++(0,0.9cm);

\node (Q11) at (0.5,-0.5) {$Q_1^{(1)}$}; 
\node (Q12) at (12.5,-0.5) {$Q_1^{(2)}$}; 

\node (Q21) at (2,-5) {$Q_2^{(1)}$};
\node (Q22) at (11,-5) {$Q_2^{(2)}$};

\node (Q31) at (4.5,-2.5) {$Q_3^{(1)}$};
\node (Q32) at (8.5,-2.5) {$Q_3^{(2)}$};  
  
\node (Q41) at (2.5,-7.75) {$Q_4^{(1)}$};
\node (Q42) at (10.5,-7.75) {$Q_4^{(2)}$};


\draw[<-, thick] (0.5,0) -- +(0,0.5) node[above] {$R_1$};

\draw[->, thick] (0.5,-1.5) to [out=270,in=105] node[right] {$f_{12}^{(1)}$} (Q21.north);
\draw[->, thick] (0.5,-1.5) to [out=225,in=125] node[left]{$f_{14}^{(1)}$} (2.25,-7.25);
\draw[->,thick] (2,-6) to [out=270,in=75] node[right,pos=0.2]{$f_{24}^{(1)}$} (Q41.north);
\draw[->,thick] (0.5,-1.5) to [out=325,in=75] node[above]{$f_{13}^{(1)}$} (Q31.north);
\draw[->,thick] (4.5,-3.5) to [out=270,in=75] node[right]{$f_{34}^{(1)}$} +(0,-2) to [out=245,in=75] (2.75,-7.25);
\draw[->,thick] (4.5,-3.5) to [out=270,in=75] node[above]{$f_{32}^{(1)}$} (2.25,-4.5);

\draw[<-, thick] (12.5,0) -- +(0,0.5) node[above] {$R_2$};

\draw[->, thick] (12.5,-1.5) to [out=270,in=75] node[left] {$f_{12}^{(2)}$} (Q22.north);

\draw[->, thick] (12.5,-1.5) to [out=305,in=55] node[right]{$f_{14}^{(2)}$} (10.75,-7.25);
\draw[->,thick] (11,-6) to [out=270,in=105] node[left,pos=0.2]{$f_{24}^{(2)}$} (Q42.north);
\draw[->,thick] (12.5,-1.5) to [out=215,in=105] node[above]{$f_{13}^{(2)}$} (Q32.north);
\draw[->,thick] (8.5,-3.5) to [out=270,in=105] node[left]{$f_{34}^{(2)}$} +(0,-2) to [out=295,in=105] (10.25,-7.25);
\draw[->,thick] (8.5,-3.5) to [out=270,in=105] node[above]{$f_{32}^{(2)}$} (10.75,-4.5);

\end{tikzpicture}
\caption{Networked system of queues.}
\label{fig:queues_new}
\vspace*{-6mm}
\end{figure}

To analyze the coding scheme, we build on the idea of tracking packets through a network of queues, as done in \cite{georgiadis2009broadcast,6522177}. 
The transmitter has two buffers, $Q_1^{(1)}$ and $Q_1^{(2)}$, to store packets destined for \Rxone, \Rxtwo, respectively.
We consider dynamic arrivals, where packets for \Rxone, \Rxtwo~arrive in each slot according to a Bernoulli process with probability $R_1$, $R_2$, respectively. These packets are called \emph{original packets}.  {An analysis for more general arrival processes is possible.}
The transmitter maintains two additional buffers, $Q_2^{(1)}$ and $Q_2^{(2)}$, for packets that were received by the wrong receiver only. 
Buffer $Q_2^{(1)}$ contains packets that are destined for \Rxone~and were received at \Rxtwo~only, and vice versa for $Q_2^{(2)}$. 
If both $Q_2^{(1)}$ and $Q_2^{(2)}$ are nonempty, the transmitter can take a packet from both queues and send their XOR combination.
Such a linear combination of original packets from $Q_2^{(1)}$ and $Q_2^{(2)}$ is called a \emph{coded} packet\footnote{The original packets arriving to $Q_1^{(j)}$ may be coded as well in the sense that error-correcting codes have been employed by lower layers. We use the term \emph{coded} packet to emphasize that information packets for different receivers have been mixed.
Alternative terms could be network-coded or inter-session-coded packet.}.
Coded packets are useful to both receivers, for each can decode a desired original packet upon reception of the coded packet.
Another necessary coding operation (see \cite[Example 2]{Kuo_Wang2014}) is as follows. The transmitter takes one packet from $Q_1^{(1)}$, e.g. $p^{(1)}$, and one packet from $Q_1^{(2)}$, e.g. $p^{(2)}$, and sends the XOR combination of the two; i.e. $p^{(1)} + p^{(2)}$.
The packets involved have not been transmitted before; hence, we call this action \emph{proactive} coding or \emph{poisoning} \cite{traskov2006network}. 
A poisoned packet is not immediately useful for any receiver. However, it becomes useful together with a \emph{remedy} packet that enables decoding of the original packet involved in the poisoned packet.  For example, assume that the poisoned packet $p^{(1)} + p^{(2)}$ was received at \Rxtwo. If, at a later stage, the corresponding \emph{remedy}
$p^{(1)}$ is  received at \Rxtwo, both $p^{(1)}$ and $p^{(2)}$ can be decoded at \Rxtwo. Therefore, upon arrival of the poisoned packet at \Rxtwo, $p^{(1)}$ becomes as useful to \Rxtwo\  as $p^{(2)}$. Since $p^{(1)}$ is desired also at \Rxone, it is  more efficient to send $p^{(1)}$ rather than $p^{(2)}$ in later channel uses; i.e., $p^{(2)}$ could, in principle, be replaced by the remedy packet $p^{(1)}$. 
\begin{remark}
If the poisoned packet is received only at \Rxj, then the remedy packet is $p^{(\bar{j})}$, for $j, \bar j \in \{1,2\}$, $j\neq \bar j$. If the poisoned packet is received at both receivers, the remedy is $p^{(1)}$ or $p^{(2)}$\!. We fix it to $p^{(1)}$. 
\end{remark}
Remedy packets are useful to both receivers. We put remedy packets into additional queues $Q_3^{(1)}$ and $Q_3^{(2)}$.
These two queues are conceptually the same queue to track the remedy packets. We draw them separately to account for correct packet arrival for each receiver separately.

The system exit for \Rxj~is represented by buffer $Q_4^{(j)}$.
Once a packet reaches $Q_4^{(j)}$, it leaves the system.
These buffers are always empty by definition.

The full networked queuing system is shown in Fig.~\ref{fig:queues_new}. This queuing system is drawn as a graphical tool to track the packets that are sent and received. We restrict the set of actions at the transmitter to 
$\mc A = \{1,2,3,4,5\}$:
\begin{itemize}
 \item$A_t=1$: send an original packet for \Rxone~from $Q_1^{(1)}$ 
 \item$A_t=2$: send an original packet for \Rxtwo~from $Q_1^{(2)}$ 
 \item$A_t=3$: send a coded packet from $Q_2^{(1)}$ and $Q_2^{(2)}$
 \item$A_t=4$: send a poisoned packet from $Q_1^{(1)}$ and $Q_1^{(2)}$
 \item$A_t=5$: send a remedy packet from $Q_3^{(1)}$ or $Q_3^{(2)}$.
\end{itemize}

\subsection{Packet Movement and Network Flow}
\label{sec:packet_movement}
As the algorithm evolves in time, packets flow on the network of Fig. \ref{fig:queues_new}.
Packet movements that are due to  \linebreak 
$A_t=1,2,3$ are based on \cite{georgiadis2009broadcast}, and are discussed in detail in \cite{georgiadis2009broadcast,6522177,heindlmaier2014oncapacity}.
Packet movements that are due to $A_t=4,5$ are based on \cite{Kuo_Wang2014}. We discuss them in our framework in the following.

Packets may move to $Q_3^{(1)}$ and $Q_3^{(2)}$ only if they are involved in a poisoned packet that is received by at least one of the receivers. To explain the packet movement from $Q_1^{(1)}$ to $Q_3^{(1)}$, for example,  consider two cases: 
\begin{itemize}
 \item[($i$)] Packet $p^{(1)}$ is a remedy packet (i.e., the poisoned packet was received at \Rxtwo). In this case, $p^{(1)}$  is moved from $Q_1^{(1)}$ to $Q_3^{(1)}$ (to be transmitted in later channel uses).
 \item[($ii$)] Packet $p^{(1)}$ is not a remedy packet (i.e. the poisoned packet was received only at \Rxone). In this case, as discussed, $p^{(2)}$ is  as useful as $p^{(1)}$ to $\Rxone$. So $p^{(1)}$ may be replaced by $p^{(2)}$ and moved from $Q_1^{(1)}$ to $Q_3^{(1)}$.
\end{itemize}

Remedy packets leave $Q_3^{(j)}$ when $A_t=5$, according to Table~\ref{tab:packet_movement}. This table appears in a similar form in \cite{Kuo_Wang2014}.

For clarification, we elaborate the case where the remedy packet sent with $A_t=5$ is received \emph{only} at \Rxone.  A remedy packet can only be sent if a poisoned packet, say $p^{(1)}+p^{(2)}$, was previously received at \Rxone~or \Rxtwo. The remedy packet is either $p^{(1)}$ or $p^{(2)}$. We consider the two cases separately: 
\begin{enumerate}
 \item 
 If $p^{(1)}$ is the remedy packet, then its arrival at \Rxone~(and not at \Rxtwo) is captured by $Q_3^{(1)}\rightarrow Q_4^{(1)}$. Furthermore, since $p^{(1)}$ is a remedy packet, it is as useful to \Rxtwo~as $p^{(2)}$. So  its arrival at \Rxone~lets us move the packet from  $Q_3^{(2)}$ to $Q_2^{(2)}$, because \Rxone~knows $p^{(1)}$. 
 \item 
 If $p^{(2)}$ is the remedy packet, then the poisoned packet $p^{(1)}+p^{(2)}$ must have been received only at \Rxone. Arrival of $p^{(2)}$ at $\Rxone$ (and not at \Rxtwo) lets \Rxone~decode $p^{(1)}$, hence $Q_3^{(1)}\rightarrow Q_4^{(1)}$. Also, since $p^{(2)}$ is received at \Rxone~and not at \Rxtwo, we have the packet movement $Q_3^{(2)}\rightarrow Q_2^{(2)}$.
\end{enumerate}

\begin{table}[t!]
\centering
 \begin{tabular}{|l|c|c|c|c|}
\hline
\!\!\!\!\parbox[t]{0.15\textwidth}{\centering Action $5$\\ (a remedy packet sent)}\!\!\!\!& \parbox[t]{0.09\textwidth}{\centering  received\\ only at \Rxone~}& \parbox[t]{0.09\textwidth}{\centering  received \\only at \Rxtwo~}& \parbox[t]{0.09\textwidth}{\centering  received\\ at \Rxone, \Rxtwo} \\ \hline \hline
Packet movement
& \parbox[t]{0.09\textwidth}{\centering $Q_3^{(1)} \!\rightarrow\! Q_4^{(1)}$\\ $Q_3^{(2)}\!\rightarrow\! Q_2^{(2)}$} 
& \parbox[t]{0.09\textwidth}{\centering$Q_3^{(1)}\!\rightarrow\! Q_2^{(1)}$\\ $Q_3^{(2)}\!\rightarrow\! Q_4^{(2)}$}
 & \parbox[t]{0.09\textwidth}{\centering $Q_3^{(1)}\!\rightarrow\! Q_4^{(1)}$\\ $Q_3^{(2)}\!\rightarrow\! Q_4^{(2)}$ } \\ \hline
 \end{tabular}
\caption{Packet movement for remedy packets in $Q_3^{(1)}$ and $Q_3^{(2)}$.}
\label{tab:packet_movement}
\vspace*{-10mm}
\end{table}

Finally, we outline two algorithms that ensure network stability for rates in $\bar{\mc C}_{\text{ fb}}^\text{mem}$: \begin{thm}
Rate-pairs $(R_1,R_2)$ are achievable if \linebreak 
$(R_1+\delta,R_2+\delta)\in\bar{\mc C}_{\text{ fb}}^\text{mem}$, for $\delta>0$.
\end{thm}

\subsection{Probabilistic Scheme}
\label{sec:probab_scheme_new}
The first scheme is a probabilistic scheme that bases its decisions on the past $L$ feedback symbols. We prove achievability of $\bar{\mc C}_{\text{ fb}}^\text{mem}(L)$ for any integer $L>0$ which approaches $\bar{\mc C}_{\text{ fb}}^\text{mem}$ exponentially fast in $L$.

Fix $L$ to be an integer and consider an encoding strategy that bases its decisions on the past $L$ feedback symbols $\ve Z^{t-1}_{t-L}$.
The decisions are random and independent from previous decisions, according to a stationary probability distribution $P_{A_t|\ve Z_{t-L}^{t-1}}(a|\ve z^L)$, $a\in\{1,\ldots, 5\}$,  $\ve z^L\in\mathcal{Z}^L$.
To simplify notation here, we write $P(a|\ve z^L)$ for $P_{A_t|\ve Z_{t-L}^{t-1}}(a|\ve z^L)$ and $P(\ve z^L)$ for $P_{\ve Z^L}(\ve z^L)$.
Based on the chosen probabilities, each link of the network in Fig.~\ref{fig:queues_new} has an effective capacity. This is the maximum  number of packets that, on average, can go through each link. Denote the capacity of the link between $Q_r^{(j)}$ and $Q_{l}^{(j)}$ by $c_{rl}^{(j)}$, where \linebreak 
$(r,l)\in\{(1,2),(1,3),(1,4),(2,4),(3,2),(3,4)\}$ and \linebreak 
$j\in \{1,2\}$. According to the packet movement in Sec.~\ref{sec:packet_movement}, we calculate the following link capacities for $j\in \{1,2\}$:
\begin{align}
\label{eq-cutvalues}
c_{12}^{(j)} &= \sum_{\ve z^L \in \mc Z^L} P(\ve z^L)   (\epsj{\ve z^L} - \epsonetwo{\ve z^L}) P(j|\ve z^L) \\
c_{13}^{(j)} &= \sum_{\ve z^L\in \mc Z^L} P(\ve z^L)   (1 - \epsonetwo{\ve z^L}) P(4|\ve z^L) \\
c_{14}^{(j)} &= \sum_{\ve z^L\in \mc Z^L} P(\ve z^L)   (1 - \epsj{\ve z^L}) P(j|\ve z^L) \\
c_{24}^{(j)} &= \sum_{\ve z^L\in \mc Z^L} P(\ve z^L)   (1 - \epsj{\ve z^L}) P(3|\ve z^L) \\
c_{32}^{(j)} &= \sum_{\ve z^L \in \mc Z^L} P(\ve z^L)   (\epsj{\ve z^L} - \epsonetwo{\ve z^L}) P(5|\ve z^L)\\
c_{34}^{(j)} &= \sum_{\ve z^L \in \mc Z^L} P(\ve z^L)   (1 - \epsj{\ve z^L}) P(5|\ve z^L). 
\end{align}

Using a max-flow analysis similar to \cite{heindlmaier2014oncapacity}, the rate pair $(R_1,R_2)$ is achievable if there exists a distribution $P_{A_t|\ve Z_{t-L}^{t-1}}$ such that the following flow optimization problem is feasible:
\begin{align}
 \label{eq:link_cap_more1}
 R_j &\!\leq\! f_{12}^{(j)} +  f_{13}^{(j)} + f_{14}^{(j)}
 \\
f_{12}^{(j)} + f_{32}^{(j)} &\!\leq\! f_{24}^{(j)}
\\
f_{13}^{(j)} &\!\leq\! f_{32}^{(j)} + f_{34}^{(j)}
\\ 
 f_{rl}^{(j)} &\!\leq\! c_{rl}^{(j)}\ \  {\forall (r,l)\!\!\in\!\!\left\{\!\!\!\!\begin{array}{l}(1,2),(1,3),(1,4)\\(2,4),(3,2),(3,4)\end{array}\!\!\!\!\right\}\!,\, \forall j\!\!\in\!\! \{1,2\}}. 
\end{align}
This is a classical flow optimization problem where link capacities can be adjusted by $P_{A_t|\ve Z_{t-L}^{t-1}}$. Note that the flow networks for \Rxone~and \Rxtwo~are  coupled only through $P_{A_t|\ve Z_{t-L}^{t-1}}$. Using the max-flow min-cut  duality theorem, the above flow problem is equivalent to the following min-cut problem:
\begin{align}
 R_j &\leq c_{12}^{(j)} +  c_{13}^{(j)} + c_{14}^{(j)}
 \label{eq:bec_cut1} \\
 R_j &\leq c_{13}^{(j)} +  c_{14}^{(j)} + c_{24}^{(j)}
 \label{eq:bec_cut2}\\
 R_j &\leq c_{12}^{(j)} +  c_{14}^{(j)} + c_{32}^{(j)} + c_{34}^{(j)}
 \label{eq:bec_cut3} \\
 R_j &\leq c_{14}^{(j)} +  c_{24}^{(j)} + c_{34}^{(j)},&\quad \forall~j\in \{1,2\}, \label{eq:bec_cut4} 
\end{align}
where \eqref{eq:bec_cut1}~-~\eqref{eq:bec_cut4} correspond to different cuts that separate $Q_1^{(j)}$ from $Q_{4}^{(j)}$, $j\in \{1,2\}$. 

The achievable rate-region is thus given by the feasibility problem defined in \eqref{eq:bec_cut1}~-~\eqref{eq:bec_cut4} over the set of probabilities $P(1|\ve z^{L}),\ldots, P(5|\ve z^{L})$, $\ve z^{L}\in\mc Z^L$.
%


\begin{prop}
\label{prop:cuts}
 Suppose $P_{A_t|\ve Z_{t-L}^{t-1}}$ is feasible for a rate pair $(R_1,R_2) \in \bar{\mc C}_{\text{ fb}}^\text{mem}(L)$.
 There exists a distribution $P^\star_{A_t|\ve Z_{t-L}^{t-1}}$ for which
\begin{itemize}
 \item $(R_1,R_2)$ is feasible and
 \item the bounds in \eqref{eq:bec_cut2} and \eqref{eq:bec_cut3} are redundant.
\end{itemize}
\end{prop}
The proof of Proposition~\ref{prop:cuts} is given in Appendix~\ref{ap-proofredundant}.
Hence, a rate pair $(R_1,R_2)$ can be achieved if there is a distribution $P_{A_t|\ve Z_{t-L}^{t-1}}$ such that
\begin{align}
 R_j \!&\leq\!\!\!\!  \sum_{\ve z^L \in \mc Z^L}\!\!\!\! P(\ve z^L) (1\!-\!\epsj{\ve z^L}\!)\!\! \left[ P(j|\ve z^L)\! +\! P(3|\ve z^L) \!+\! P(5|\ve z^L) \right] \label{eq:bec_more_constr1}\\
 R_j\! &\leq\!\!\!\!  \sum_{\ve z^L \in \mc Z^L}\!\!\!\! P(\ve z^L) (1\!-\!\epsonetwo{\ve z^L}\!)\!\! \left[ P(j|\ve z^L) \!+\! P(4|\ve z^L) \right]. \label{eq:bec_more_constr2}
\end{align}

One can verify that the region described by \eqref{eq:bec_more_constr1}~-~\eqref{eq:bec_more_constr2} is equivalent to $\bar{\mc C}_{\text{ fb}}^\text{mem}(L)$ by choosing a feasible set of distributions $P(1|\ve z^{L}),\ldots, P(5|\ve z^{L})$, $\ve z^{L}\in\mc Z^L$ such that
\begin{align}
P(1|\ve z^L)\! +\! P(3|\ve z^L) \!+\! P(5|\ve z^L)&= x(\ve z^L),\quad \forall \ve z^L\in\mc Z^L \label{eq:proba_relation_xs} \\
P(2|\ve z^L)\! +\! P(3|\ve z^L) \!+\! P(5|\ve z^L)&= y(\ve z^L),\quad \forall \ve z^L\in\mc Z^L. \label{eq:proba_relation_ys}
\end{align}

\begin{remark}
 Note that the mapping from $x(\ve z^L)$, $y(\ve z^L)$,\linebreak  $\ve z^L\in\mc Z^L$ to $P_{A_t|\ve Z^{L}}$ defined in \eqref{eq:proba_relation_xs}~-~\eqref{eq:proba_relation_ys} is not unique. In particular, for some valid choices of $P_{A_t|\ve Z^{L}}$ the constraints in \eqref{eq:bec_cut2} and \eqref{eq:bec_cut3} will \emph{not} be redundant. Nevertheless, Proposition~\ref{prop:cuts} tells us that one can always find a distribution $P^\star_{A_t|\ve Z^{L}}$ 
\begin{itemize}
 \item that satisfies \eqref{eq:proba_relation_xs}~-~\eqref{eq:proba_relation_ys} and
 \item for which \eqref{eq:bec_cut2} and \eqref{eq:bec_cut3} are redundant,
\end{itemize}
for any values of $x(\ve z^L)$, $y(\ve z^L)$ $\ve z^L\in\mc Z^L$.
For a given set of $x(\ve z^L)$, $y(\ve z^L)$ $\ve z^L\in\mc Z^L$, we first choose a distribution $P_{A_t|\ve Z^{L}}$ satisfying \eqref{eq:proba_relation_xs}~-~\eqref{eq:proba_relation_ys} and then determine $P_{A_t|\ve Z^{L}}^\star$ from $P_{A_t|\ve Z^{L}}$.
\end{remark}

This shows that a feasible set of distributions $P(1|\ve z^{L}),\ldots, P(5|\ve z^{L})$, $\ve z^{L}\in\mc Z^L$ can be found if $(R_1,R_2)\in \bar{\mc C}_{\text{ fb}}^\text{mem}(L)$ and hence $\bar{\mc C}_{\text{ fb}}^\text{mem}(L)$ is achievable.


\subsection{Deterministic Scheme}
\label{sec:new_ach_det}
For the probabilistic scheme, one must compute the optimal set of probability distributions in advance. This set depends on $R_1$, $R_2$ which might be unknown to the transmitter ahead of time. Furthermore, in the probabilistic approach, it might happen that an action is chosen but there is no packet in the corresponding queues to be transmitted.
We avoid both drawbacks by a max-weight backpressure-like algorithm \cite{tassiulas1992stability,neely2010stochastic} that bases its actions on both the queue state and the feedback. 
In each slot $t$,
a weight function is computed for each action and the action with the higher weight is chosen.
Table~\ref{tab:new_det_algo_hmm} lists the weights for each action depending on the current queue state $\ve Q_t$ and the previous feedback state $\ve Z^{t-1}$. 
The queue state $\ve Q_t$ is a vector with one entry for each queue in the network
containing the number of packets of each queue at time $t$.
Note that the values of $\epsone{\ve z^{t-1}}$, $\epstwo{\ve z^{t-1}}$ and $\epsonetwo{\ve z^{t-1}}$ can be computed recursively. 

\begin{table}[t!]
\centering
 \begin{tabular}{|c|l|}
 \hline
 $A_t$ & Weight depending on queue lengths and $\ve Z^{t-1}=\ve z^{t-1}$  \\ \hline
 $1$ &  $[1-\epsone{\ve z^{t-1}}] Q_{1,t}^{(1)} + \epsonenottwo{\ve z^{t-1}}  (Q_{1,t}^{(1)}-Q_{2,t}^{(1)}) $\\
 $2$ &  $[1-\epstwo{\ve z^{t-1}}] Q_{1,t}^{(2)} + \epsnotonetwo{\ve z^{t-1}}  (Q_{1,t}^{(2)}-Q_{2,t}^{(2)}) $\\
 $3$ &  $[1- \epsone{\ve z^{t-1}}]  Q_{2,t}^{(1)} + [1- \epstwo{\ve z^{t-1}}]  Q_{2,t}^{(2)}$\\
 $4$ &  $[1- \epsonetwo{\ve z^{t-1}}]  \left(  Q_{1,t}^{(1)} -  Q_{3,t}^{(1)} + Q_{1,t}^{(2)} -  Q_{3,t}^{(2)}  \right) $\\
 $5$ & $\epsonenottwo{\ve z^{t-1}}  (Q_{3,t}^{(1)}-Q_{2,t}^{(1)})  + [1-\epsone{\ve z^{t-1}}] Q_{3,t}^{(1)}$\\ &$+ \epsnotonetwo{\ve z^{t-1}}  (Q_{3,t}^{(2)}-Q_{2,t}^{(2)})  + [1-\epstwo{\ve z^{t-1}}] Q_{3,t}^{(2)}   $\\
\hline
 \end{tabular}
\caption{Deterministic scheme with $A_t \in \mc A$. $Q_{l,t}^{(j)}$ denotes the number of packets in queue $Q_{l}^{(j)}$ at time $t$.}
\label{tab:new_det_algo_hmm}
\vspace*{-8mm}
\end{table}

\begin{prop}
The strategy in 
Table~\ref{tab:new_det_algo_hmm}
stabilizes all queues in the network for every $(R_1+\delta, R_2+\delta) \in \bar{\mc C}_{\text{ fb}}^\text{mem}$, $\delta>0$.
\label{prop:maxweight_new_hmm}
\end{prop}
Hence, the strategy in Table~\ref{tab:new_det_algo_hmm} provides an implicit way to compute the capacity region.
The proof of Prop.~\ref{prop:maxweight_new_hmm} in Appendix~\ref{ap-proof_prop_maxweight_hmm} 
uses a finite-horizon Lyapunov drift analysis \cite{neely2010stochastic} that is adapted to account for \emph{previous} observations (rather than the current channel state). 

\section{Conclusion}
\label{sec:results}

We studied the two-receiver broadcast packet erasure channel with feedback and hidden channel memory. The channel memory evolves according to a Markov chain that is not observable by the transmitter. We developed outer bounds that are achievable with two outlined coding schemes.

\section*{Acknowledgments}
The authors are supported by the German Ministry of Education and Research through an Alexander von Humboldt-Professorship, by the grant DLR@Uni of the Helmholtz Allianz and by the Swiss National Science Foundation fellowship no. 146617.
The authors thank Gerhard Kramer and the anonymous reviewers for helpful comments.

\newpage
\appendices
\section{Proof of Theorem \ref{outerbound}}
\label{ap-outerboundproof}
\begin{figure}[t]
\centering
  \tikzset{>=latex}
\begin{tikzpicture}[scale = 0.5, font=\small]
\input{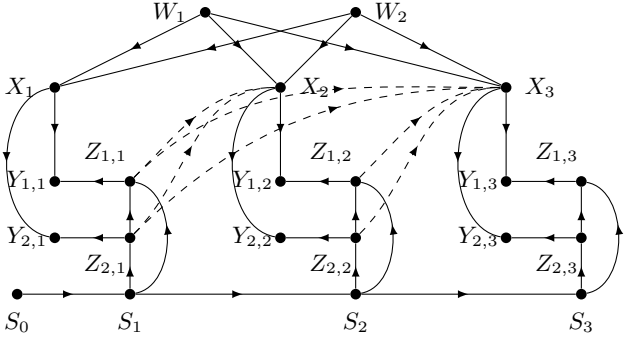}
 \end{tikzpicture}
\caption{Bayesian network for the two-receiver broadcast packet erasure channel with memory and ACK/NACK  feedback, for $n=3$ and $d=1$. Dependencies due to feedback are drawn with dashed lines.}
\label{fig:fdg_system}
\end{figure}

Theorem \ref{outerbound} is similar to the outer bound in \cite{heindlmaier2014oncapacity}. We outline the proof here.
The joint probability mass function of the system then factorizes as
\begin{align}
&P_{W_1 W_2 X^n S^n Y_1^n Y_2^n Z_1^n Z_2^n}= 
P_{W_1} P_{W_2} P_{S_0} \prod_{t=1}^n P_{S_t|S_{t-1}}\cdot \nonumber \\
&\cdot  P_{X_t|Z_1^{t-1} Z_2^{t-1}} P_{Z_{2,t}|S_t} P_{Z_{1,t}|Z_{2,t}S_t} P_{Y_{1,t}|X_t Z_{1,t}} P_{Y_{2,t}|X_t Z_{2,t}}. \nonumber
\end{align}
The corresponding Bayesian network is shown in Fig.~\ref{fig:fdg_system}.
We write $ P_{\ve Z^{t\hspace{-.025cm}-\hspace{-.025cm}1}}(\ve z^{t\hspace{-.025cm}-\hspace{-.025cm}1})$ as $ P(\ve z^{t\hspace{-.025cm}-\hspace{-.025cm}1})$ and bound
\allowdisplaybreaks
\begin{align}
&R_1\!-\!\delta\leq \frac{1}{n}I(W_1;Y_1^n)\nonumber \\
&\leq \frac{1}{n}I(W_1;Y_1^n Z_2^{n-1})\nonumber \\
&=\frac{1}{n}\sum_{t=1}^n I(W_1;Y_{1,t} Z_{2,t\hspace{-.025cm}-\hspace{-.025cm}1} |Y_1^{t\hspace{-.025cm}-\hspace{-.025cm}1} Z_2^{t-2} )\nonumber \\
&=\frac{1}{n}\sum_{t=1}^n \left[I(W_1; Z_{2,t\hspace{-.025cm}-\hspace{-.025cm}1}|Y_1^{t\hspace{-.025cm}-\hspace{-.025cm}1}Z_1^{t\hspace{-.025cm}-\hspace{-.025cm}1}Z_2^{t-2})\right.\nonumber\\
&\quad\quad\quad\quad\left.+ I(W_1;Y_{1,t}|Y_1^{t\hspace{-.025cm}-\hspace{-.025cm}1}Z_2^{t\hspace{-.025cm}-\hspace{-.025cm}1})\right]\nonumber \\
&\stackrel{(a)}{=}\sum_{t=1}^n \frac{1}{n}I(W_1;Y_{1,t}|Y_1^{t\hspace{-.025cm}-\hspace{-.025cm}1}Z_1^{t\hspace{-.025cm}-\hspace{-.025cm}1}Z_2^{t\hspace{-.025cm}-\hspace{-.025cm}1})\nonumber \\
&=\frac{1}{n}\! \sum_{t=1}^n \sum_{\ve z^{t\hspace{-.025cm}-\hspace{-.025cm}1} } \!P(\ve z^{t\hspace{-.025cm}-\hspace{-.025cm}1}) I(W_1;Y_{1,t}|Y_1^{t\hspace{-.025cm}-\hspace{-.025cm}1}\!,\ve Z^{t\hspace{-.025cm}-\hspace{-.025cm}1}\!=\!\ve z^{t\hspace{-.025cm}-\hspace{-.025cm}1}).\label{no1_hmm}
\end{align}
In the above chain of inequalities, $(a)$ follows because $Z_1^{t\hspace{-.025cm}-\hspace{-.025cm}1}$ is a function of $Y_1^{t\hspace{-.025cm}-\hspace{-.025cm}1}$ and 
 $W_1-Y_1^{t\hspace{-.025cm}-\hspace{-.025cm}1}Z_1^{t\hspace{-.025cm}-\hspace{-.025cm}1}Z_2^{t-2}- Z_{2,t\hspace{-.025cm}-\hspace{-.025cm}1}$ forms a Markov chain for every $t=1,\ldots,n$.
Markov chains can be checked in Fig.~\ref{fig:fdg_system}.
Similarly, we write 
\begin{align}
&R_1\!-\!\delta\!\leq\! \frac{1}{n}I(W_1;Y_1^nY_2^n|W_2)\nonumber\\
&\!=\!\frac{1}{n} \sum_{t=1}^n I(W_1;Y_{1,t}Y_{2,t}|W_2 Y_{1}^{t\hspace{-.025cm}-\hspace{-.025cm}1}Y_{2}^{t\hspace{-.025cm}-\hspace{-.025cm}1})\nonumber\\
&\!\stackrel{(a)}{=}\!\frac{1}{n} \!\sum_{t=1}^n \!\sum_{\ve z^{t\hspace{-.025cm}-\hspace{-.025cm}1} } P(\hspace{-.025cm}\ve z^{t\hspace{-.025cm}-\hspace{-.025cm}1}\hspace{-.025cm}) I(\hspace{-.05cm}W_1;Y_{1,t}Y_{2,t}|W_2 Y_{1}^{t\hspace{-.025cm}-\hspace{-.025cm}1}Y_{2}^{t\hspace{-.025cm}-\hspace{-.025cm}1}\!, \ve Z^{t\hspace{-.025cm}-\hspace{-.025cm}1}\!\!=\!\hspace{-.025cm}\ve z^{t\hspace{-.025cm}-\hspace{-.025cm}1}\hspace{-.05cm})\nonumber.
\end{align}
where $(a)$
follows because $\ve Z^{t\hspace{-.025cm}-\hspace{-.025cm}1}$ is a function of $Y_1^{t\hspace{-.025cm}-\hspace{-.025cm}1},Y_2^{t\hspace{-.025cm}-\hspace{-.025cm}1}$.
We next obtain a relationship using a derivation similar to \cite[Lemma 1]{heindlmaier2014oncapacity}. For every $t=1,\ldots, n$, $\ve z^{t\hspace{-.025cm}-\hspace{-.025cm}1}\in \mc Z^{t\hspace{-.025cm}-\hspace{-.025cm}1}$, $j=1,2$, we have
\begin{align*}
&I(W_1;Y_{1,t}|Y_1^{t\hspace{-.025cm}-\hspace{-.025cm}1},\ve Z^{t\hspace{-.025cm}-\hspace{-.025cm}1}\!=\!\ve z^{t\hspace{-.025cm}-\hspace{-.025cm}1})\nonumber \\
&\quad=(1-\epsj{\ve z^{t\hspace{-.025cm}-\hspace{-.025cm}1}}) I(W_1;X_t|Y_1^{t\hspace{-.025cm}-\hspace{-.025cm}1},\ve Z^{t\hspace{-.025cm}-\hspace{-.025cm}1}\!=\!\ve z^{t\hspace{-.025cm}-\hspace{-.025cm}1}),\\
&I(W_1;Y_{1,t}Y_{2,t}|W_2 Y_{1}^{t\hspace{-.025cm}-\hspace{-.025cm}1}Y_{2}^{t\hspace{-.025cm}-\hspace{-.025cm}1}, \ve Z^{t\hspace{-.025cm}-\hspace{-.025cm}1}\!=\!\ve z^{t\hspace{-.025cm}-\hspace{-.025cm}1})\nonumber \\
&\quad = (1-\epsonetwo{\ve z^{t\hspace{-.025cm}-\hspace{-.025cm}1}}) I(W_1;X_t|W_2 Y_{1}^{t\hspace{-.025cm}-\hspace{-.025cm}1}Y_{2}^{t\hspace{-.025cm}-\hspace{-.025cm}1}, \ve Z^{t\hspace{-.025cm}-\hspace{-.025cm}1}\!=\!\ve z^{t\hspace{-.025cm}-\hspace{-.025cm}1}).
\end{align*}
Define the variables:
\begin{align}
\label{definition}
\begin{array}{l} u^{(j)}(\ve z^{t\hspace{-.025cm}-\hspace{-.025cm}1})\! =\! I(W_1;X_{t}|Y_1^{t\hspace{-.025cm}-\hspace{-.025cm}1}, \ve Z^{t\hspace{-.025cm}-\hspace{-.025cm}1}\!=\!\ve z^{t\hspace{-.025cm}-\hspace{-.025cm}1}) \\
 v^{(j)}(\ve z^{t\hspace{-.025cm}-\hspace{-.025cm}1})  \!=\! I(W_1;X_{t}|W_2 Y_1^{t\hspace{-.025cm}-\hspace{-.025cm}1} Y_2^{t\hspace{-.025cm}-\hspace{-.025cm}1}, \ve Z^{t\hspace{-.025cm}-\hspace{-.025cm}1}\!=\!\ve z^{t\hspace{-.025cm}-\hspace{-.025cm}1})
 \end{array}
\end{align}
From \cite[Lemma 2]{heindlmaier2014oncapacity}, for every $t=1,\ldots,n$, $j = 1,2$, \linebreak 
$\ve z^{t\hspace{-.025cm}-\hspace{-.025cm}1}\in\mc Z^{t\hspace{-.025cm}-\hspace{-.025cm}1}$, we have
\begin{align}
 u^{(j)}(\ve z^{t\hspace{-.025cm}-\hspace{-.025cm}1}) +  v^{(\bar j)}(\ve z^{t\hspace{-.025cm}-\hspace{-.025cm}1}) \leq 1.
\end{align}
We thus obtain the following feasibility problem as an outer bound on the achievable rate-region:
\begin{align}
0\leq &u^{(j)}(\ve z^{t\hspace{-.025cm}-\hspace{-.025cm}1}),  v^{(j)}(\ve z^{t\hspace{-.025cm}-\hspace{-.025cm}1})\leq 1 \qquad~ ~ \forall~t, \ve z^{t\hspace{-.025cm}-\hspace{-.025cm}1}\\
& u^{(j)}(\ve z^{t\hspace{-.025cm}-\hspace{-.025cm}1})+v^{(\bar{j})}(\ve z^{t\hspace{-.025cm}-\hspace{-.025cm}1})\leq 1 \qquad \forall~t, \ve z^{t\hspace{-.025cm}-\hspace{-.025cm}1} \label{eq:ob_ineq} \\
R_j\leq & \frac{1}{n} \sum_{t=1}^n \sum_{\ve z^{t\hspace{-.025cm}-\hspace{-.025cm}1}} P(\ve z^{t\hspace{-.025cm}-\hspace{-.025cm}1}) (1-\epsj{\ve z^{t\hspace{-.025cm}-\hspace{-.025cm}1}}) u^{(j)}(\ve z^{t\hspace{-.025cm}-\hspace{-.025cm}1}) \\
R_j\leq & \frac{1}{n} \sum_{t=1}^n \sum_{\ve z^{t\hspace{-.025cm}-\hspace{-.025cm}1}} P(\ve z^{t\hspace{-.025cm}-\hspace{-.025cm}1}) (1-\epsonetwo{\ve z^{t\hspace{-.025cm}-\hspace{-.025cm}1}}) v^{(j)}(\ve z^{t\hspace{-.025cm}-\hspace{-.025cm}1})
\end{align}
The inequality \eqref{eq:ob_ineq} can be made tight without changing the rate region. By setting $u^{(1)}(\ve z^L) = x(\ve z^L)$ and \linebreak 
$u^{(2)}(\ve z^L) = y(\ve z^L)$,
the above characterization can be written as in \eqref{eq:posouter_hmm}~-~\eqref{eq:R2_constr2outer_hmm}.

\section{Derivation of Corollary~\ref{cor:Pz_decay}}
\label{sec:derivation_corollary_forgetting}
The corollary follows from Theorem~\ref{thm:hmm_forgetting} and the following derivation:
\begin{align}
  &\sum_{\ve z_{t}} \left| P(\ve z_{t}|\ve z^{t-1}) -  P(\ve z_{t}|\ve z^{t-1}_{t-L})    \right| \nonumber\\
  =&\sum_{\ve z_{t}} \left| \sum_{s_t} \left( P(\ve z_{t}, s_t|\ve z^{t-1}) -  P(\ve z_{t}, s_t|\ve z^{t-1}_{t-L}) \right)   \right| \nonumber \\
  =&\sum_{\ve z_{t}} \left| \sum_{s_t} \left( P(s_t|\ve z^{t-1}) -  P(s_t|\ve z^{t-1}_{t-L}) \right)  P(\ve z_t|s_t) \right| \nonumber \\
  \leq&  \sum_{s_t} \left| P(s_t|\ve z^{t-1}) -  P(s_t|\ve z^{t-1}_{t-L}) \right| \sum_{\ve z_{t}} P(\ve z_t|s_t) \nonumber \\
  =&\sum_{s_t} \left| P(s_t|\ve z^{t-1}) -  P(s_t|\ve z^{t-1}_{t-L}) \right|
\end{align}

\section{Approximation of Outer Bounds}
\label{sec:proof_bound_hidden_approx}
We derive only the first inequality for the outer approximation of the outer bound in \eqref{eq:R1_constr1outer_hmm_wind_outer}. The other inequalities follow similarly. We write $P_{\ve Z^{t\hspace{-.025cm}-\hspace{-.025cm}1}}(\ve z^{t\hspace{-.025cm}-\hspace{-.025cm}1})$ as $P(\ve z^{t\hspace{-.025cm}-\hspace{-.025cm}1})$ and bound
\begin{align}
R_1\!\stackrel{(a)}{\leq}\!& \frac{1}{n} \sum_{t=1}^n \sum_{\ve z^{t\hspace{-.025cm}-\hspace{-.025cm}1}} \left(\begin{array}{l}P(\ve z^{t\hspace{-.025cm}-\hspace{-.025cm}1}) (1-\epsone{\ve z^{t\hspace{-.025cm}-\hspace{-.025cm}1}}) \\\quad\times I(W_1;X_{t}|Y_1^{t\hspace{-.025cm}-\hspace{-.025cm}1}, \ve Z^{t\hspace{-.025cm}-\hspace{-.025cm}1}=\ve z^{t\hspace{-.025cm}-\hspace{-.025cm}1})\end{array}\right)\nonumber \\
\leq &\frac{1}{n} \sum_{t=1}^n \sum_{\ve z^{t\hspace{-.025cm}-\hspace{-.025cm}1}} \left(\!\!\!\!\begin{array}{l}P(\ve z^{t\hspace{-.025cm}-\hspace{-.025cm}1}) \left( (1-\epsone{\ve z^{t\hspace{-.025cm}-\hspace{-.025cm}1}_{t-L}} + 2  (1-\sigma)^L \right)\\ \quad \times  I(W_1;X_{t}|Y_1^{t\hspace{-.025cm}-\hspace{-.025cm}1}, \ve Z^{t\hspace{-.025cm}-\hspace{-.025cm}1}=\ve z^{t\hspace{-.025cm}-\hspace{-.025cm}1})\end{array}\!\!\!\!\right) \nonumber \\
\leq &\frac{1}{n} \sum_{t=1}^n \!\!\left[\!\! \left( \sum_{\ve z^{t\hspace{-.025cm}-\hspace{-.025cm}1}} \begin{array}{l}P(\ve z^{t\hspace{-.025cm}-\hspace{-.025cm}1}) (1\!-\!\epsone{\ve z^{t\hspace{-.025cm}-\hspace{-.025cm}1}_{t-L}} )\\ \quad\times I(W_1;X_{t}|Y_1^{t\hspace{-.025cm}-\hspace{-.025cm}1}\!, \ve Z^{t\hspace{-.025cm}-\hspace{-.025cm}1}\!=\!\ve z^{t\hspace{-.025cm}-\hspace{-.025cm}1})\end{array}\!\! \right) \right. \nonumber\\
&\quad\qquad  + 2  (1-\sigma)^L  I(W_1;X_{t}|Y_1^{t\hspace{-.025cm}-\hspace{-.025cm}1}, \ve Z^{t\hspace{-.025cm}-\hspace{-.025cm}1})  \Bigg]\nonumber \\
\leq&  \frac{1}{n} \sum_{t=1}^n\! \sum_{\ve z^{t\hspace{-.025cm}-\hspace{-.025cm}1}_{t-L}}\!\! \left(\!\!\!\!\begin{array}{l}P_{\ve Z^{t\hspace{-.025cm}-\hspace{-.025cm}1}_{t-L}}(\ve z^{t\hspace{-.025cm}-\hspace{-.025cm}1}_{t-L}) \left(1 - \epsone{\ve z^{t\hspace{-.025cm}-\hspace{-.025cm}1}_{t-L}}\right)  \\\quad \times I(W_1;X_{t}|Y_1^{t\hspace{-.025cm}-\hspace{-.025cm}1} \ve Z^{t-L-1},  \ve Z^{t\hspace{-.025cm}-\hspace{-.025cm}1}_{t-L}\!=\!\ve z^{t\hspace{-.025cm}-\hspace{-.025cm}1}_{t-L})\end{array}\!\!\!\!\right) \nonumber\\
\nonumber&+2 (1-\sigma)^L ,
\end{align}
where $(a)$ follows from \eqref{no1_hmm}.
The sequence $\ve Z^n$ is stationary, i.e. $P_{\ve Z^{t\hspace{-.025cm}-\hspace{-.025cm}1}_{t-L}}(\ve z^L)$ does not depend on $t$ but only on $\ve z^L$.
Hence one can simplify the above equations using a time-sharing argument with time-sharing random variable $T$:
\begin{align}
 R_1
 \leq&  \sum_{\ve z^L}  \!\! \left(\!\!\!\!\begin{array}{l}P_{\ve Z^L}(\ve z^L) \left(1-\epsone{\ve z^L} \right)  \\\quad \times I(W_1;X_{T}|T Y_1^{T-1} \ve Z^{T-L-1},  \ve Z^{T-1}_{T-L}\!=\!\ve z^L) \end{array}\!\!\!\!\right) \nonumber\\
&+2 (1-\sigma)^L 
\end{align}
Define
\begin{align}
 x(\ve z^L) & = I(W_1;X_{T}|T Y_1^{T\hspace{-.025cm}-\hspace{-.025cm}1} \ve Z^{T-L-1},  \ve Z^{T\hspace{-.025cm}-\hspace{-.025cm}1}_{T-L}=\ve z^L). \nonumber
\end{align}
We obtain the outer approximation of the outer bound in \eqref{eq:R1_constr1outer_hmm_wind_outer}:
\begin{align}
R_1
\leq 2  (1-\sigma)^L + \sum_{\ve z^L} P_{\ve Z^L}(\ve z^L) (1-\epsone{\ve z^L}) x(\ve z^L).
\end{align}
An inner approximation can similarly be derived with a negative sign for $2  (1-\sigma)^L$. The outer bound is sandwiched between the inner and outer approximation.

\section{Proof of Proposition~\ref{prop:cuts}}
\label{ap-proofredundant}

\begin{figure}[ht]
\centering
\begin{tikzpicture}[scale=0.7]
\node (Q11) at (0.5,-0.5) {$Q_1^{(j)}$}; 
\node (Q21) at (-1,-4.5) {$Q_2^{(j)}$}; 
\node (Q31) at (3,-3.5) {$Q_3^{(j)}$};
\node (Q41) at (1,-8.5) {$Q_4^{(j)}$};

\draw[->, thick] (Q11) to [out=270,in=80] node[right] {$c_{12}^{(j)}$} (Q21.north);
\draw[->, thick] (Q11) to [out=225,in=65] node[left]{$c_{14}^{(j)}$} +(-3,-3) to [out=245,in=125] (Q41);
\draw[->,thick] (Q21) to [out=270,in=105] node[right,pos=0.2]{$c_{24}^{(j)}$} (Q41);
\draw[->,thick] (Q11) to [out=325,in=75] node[above]{$c_{13}$} (Q31.north);
\draw[->,thick] (Q31) to [out=270,in=75] node[right]{$c_{34}^{(j)}$} +(0,-2) to [out=245,in=75] (Q41);
\draw[->,thick] (Q31) to [out=190,in=25] node[below]{$c_{32}^{(j)}$} (Q21.east);

\node (A) at (-1,-0.5) {$A_j$};
\draw[dashed] (A) to [out=290,in=270] (1.75,-0.5); 

\node (C) at (4,-7) {$C_j$};
\draw[dashed] (-2,-0.5) to (C);

\node (B) at (-3,-7) {$B_j$};
\draw[dashed] (5,-0.5) to (B);

\node (D) at (3,-8) {$D_j$};
\draw[dashed] (-1,-8) to [out=80,in=110] (D);
\end{tikzpicture}
\caption{Queue network and cuts.}
\label{fig:queues_cuts}
\end{figure}
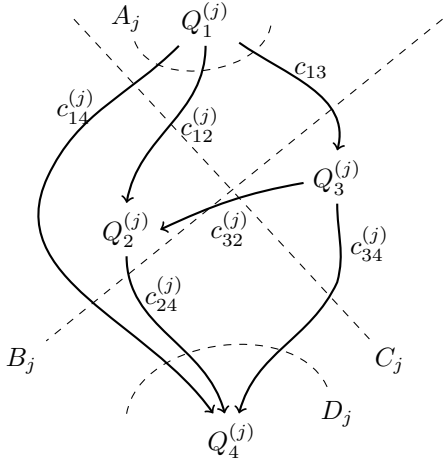

We again write $P(a|\ve z^L)$ for $P_{A_t|\ve Z_{t-L}^{t-1}}(a|\ve z^L)$ and define the variables $A_j$, $B_j$, $C_j$, $D_j$, $j \in \{1,2\}$ as the cut values defined in \eqref{eq:bec_cut1}~-~\eqref{eq:bec_cut4} and shown in Fig.~\ref{fig:queues_cuts}:
\begin{align}
 A_j &= c_{12}^{(j)} +  c_{13} + c_{14}^{(j)} \nonumber\\
     &= \hspace*{-.5em}\sum_{\ve z^L \in \mc Z^L} \hspace*{-.5em} P(\ve z^L)  (1-\epsonetwo{\ve z^L}) \left[ P(j|\ve z^L) + P(4|\ve z^L) \right] \nonumber \\
 B_j &= c_{13} +  c_{14}^{(j)} + c_{24}^{(j)} \nonumber\\
     &= \hspace*{-.5em}\sum_{\ve z^L \in \mc Z^L} \hspace*{-.5em} P(\ve z^L) \Big[ (1-\epsonetwo{\ve z^L})  P(4|\ve z^L)  \nonumber \\
     & \qquad \qquad \qquad  + (1-\epsj{\ve z^L}) \left[ P(j|\ve z^L) + P(3|\ve z^L) \right]  \Big] \nonumber \\
 C_j &= c_{12}^{(j)} +  c_{14}^{(j)} + c_{32}^{(j)} + c_{34}^{(j)} \nonumber\\
     &= \hspace*{-.5em}\sum_{\ve z^L \in \mc Z^L} \hspace*{-.5em} P(\ve z^L)  (1-\epsonetwo{\ve z^L}) \left[ P(j|\ve z^L) + P(5|\ve z^L) \right] \nonumber \\
 D_j &= c_{14}^{(j)} +  c_{24}^{(j)} + c_{34}^{(j)} \nonumber\\
     &= \hspace*{-.5em}\sum_{\ve z^L \in \mc Z^L} \hspace*{-.5em} P(\ve z^L)  
      (1\!-\!\epsj{\ve z^L}\!)\!\! \left[ P(j|\ve z^L)\! +\! P(3|\ve z^L) \!+\! P(5|\ve z^L) \right] \nonumber
\end{align}
We omit the superscript $(j)$ for $c_{13}$ to emphasize that\linebreak $c_{13}^{(1)}= c_{13}^{(2)}= c_{13} $.

Our goal is to show that for any link capacities $c_{rl}^{(j)}$ and associated cut values $A_j$, $B_j$, $C_j$, $D_j$ induced by a distribution $P_{A_t|\ve Z_{t-L}^{t-1}}$, there is another distribution $P^\star_{A_t|\ve Z_{t-L}^{t-1}}$ with associated link capacities $c_{rl}^{(j)\star}$ and cut values $A^\star_j$, $B^\star_j$, $C^\star_j$, $D^\star_j$ such that 
\begin{align}
 \min\left\{ A^\star_j, B^\star_j, C^\star_j, D^\star_j \right\} &= \min\left\{ A_j, B_j, C_j, D_j \right\} \label{eq:cuts_criterion1} \\
 \min\left\{ A^\star_j, B^\star_j, C^\star_j, D^\star_j \right\} &= \min\left\{ A^\star_j, D^\star_j \right\}. \label{eq:cuts_criterion2}
\end{align}
That is, the minimal cut value does not change under $P^\star_{A_t|\ve Z_{t-L}^{t-1}}$, but the cuts $B^\star_j$ and $C^\star_j$ are redundant.
The next lemma states one sufficient condition for redundancy:
\begin{lemma}
\label{lem:flow_div_equality}
 The cut values $B_j$ and $C_j$ are redundant for both $j=1,2$ if $P_{A_t|\ve Z_{t-L}^{t-1}}$ satisfies
 \begin{align}
c_{13} &= c_{32}^{(j)} + c_{34}^{(j)}\quad \forall~j\in \{1,2\}.  \label{eq:flow_div_equality} 
\end{align}
\end{lemma}
\begin{IEEEproof}
The condition in \eqref{eq:flow_div_equality} is equivalent to 
\begin{align}
\sum_{\ve z^L \in \mc Z^L} P(\ve z^L)   (1 - \epsonetwo{\ve z^L}) \left[ P(4|\ve z^L) - P(5|\ve z^L) \right] & =  0. 
\end{align}
 One can verify that $C_j = A_j$ in this case, so $C_j$ can be omitted.
Additionally, $D_j \leq B_j$ because
\begin{align}
 &B_j - D_j= \nonumber\\
 =&c_{13} +  c_{14}^{(j)} + c_{24}^{(j)} - \left( c_{14}^{(j)} +  c_{24}^{(j)} + c_{34}^{(j)} \right) \nonumber \\
=&\hspace*{-.5em} \sum_{\ve z^L \in \mc Z^L} \hspace*{-.5em}P(\ve z^L)  \left( (1 - \epsonetwo{\ve z^L}) P(4|\ve z^L) - (1 - \epsj{\ve z^L}) P(5|\ve z^L) \right) \nonumber \\
=&\hspace*{-.5em} \sum_{\ve z^L \in \mc Z^L} \hspace*{-.5em}P(\ve z^L)  \left( (1 - \epsonetwo{\ve z^L}) P(5|\ve z^L) - (1 - \epsj{\ve z^L}) P(5|\ve z^L) \right) \nonumber \\ 
\geq & 0, 
\end{align}
where the last inequality is due to $1 - \epsonetwo{\ve z^L} \geq 1 - \epsj{\ve z^L}$ for all $\ve z^L \in \mc Z^L$ and $j=\{1,2\}$.
\end{IEEEproof}

\begin{remark}
 Note that one \emph{cannot} simply set $P^\star(4|\ve z^L) = P^\star(5|\ve z^L)$, $\forall \ve z^L \in \mc Z^L$, to ensure \eqref{eq:flow_div_equality}.
 By the mapping in \eqref{eq:proba_relation_xs}~-~\eqref{eq:proba_relation_ys}, this would imply $x(\ve z^L) + y(\ve z^L) \geq 1$, $\forall \ve z^L \in \mc Z^L$ and hence permit only a limited set of values for $x(\ve z^L)$ and $y(\ve z^L)$. 
 This additional constraint leads to a smaller rate region, as derived in \cite{heindlmaier2014oncapacity}.
\end{remark}

%
For any given distribution $P_{A_t|\ve Z_{t-L}^{t-1}}$, we choose $P^\star_{A_t|\ve Z_{t-L}^{t-1}}$ such that $\forall \ve z^L \in \mc Z^L$:
\begin{align}
 P^\star(1|\ve z^L) &= P(1|\ve z^L) \label{eq:pstar1} \\
 P^\star(2|\ve z^L) &= P(2|\ve z^L) \\ 
 P^\star(4|\ve z^L) &= P(4|\ve z^L) \\
 P^\star(3|\ve z^L)+ P^\star(5|\ve z^L) & = P(3|\ve z^L)+ P(5|\ve z^L) \label{eq:sum_constr}
\end{align}
By choosing $P^\star(3|\ve z^L) \neq P(3|\ve z^L)$, $P^\star(5|\ve z^L) \neq P(5|\ve z^L)$, the link capacities $c_{24}^{(j)\star}$, $c_{34}^{(j)\star}$ and $c_{32}^{(j)\star}$, $j \in \{1,2\}$ are varied. 
The other link capacities are not affected, hence 
\begin{align}
 c_{12}^{(j)\star} = c_{12}^{(j)}, \quad
 c_{13}^{(j)\star} = c_{13}^{(j)}, \quad
 c_{14}^{(j)\star} = c_{14}^{(j)}, ~ j \in \{1,2\}. 
\end{align}
In the following we therefore omit the superscript $\star$ for those capacities that stay constant with $P_{A_t|\ve Z_{t-L}^{t-1}}$ and $P^\star_{A_t|\ve Z_{t-L}^{t-1}}$.
Note that $A^\star_j=A_j$, $D^\star_j=D_j$, $j=\{1,2\}$
because 
\begin{itemize}
 \item $A^\star_j$ is not affected by changing $P^\star(3|\ve z^L)$ or $P^\star(5|\ve z^L)$ and
 \item $D^\star_j$ depends only on the sum $P^\star(3|\ve z^L)+P^\star(5|\ve z^L)$ that is kept constant in \eqref{eq:sum_constr}, hence
 \begin{align}
  c_{24}^{(j)} + c_{34}^{(j)} &= c_{24}^{(j)\star} + c_{34}^{(j)\star}. \label{eq:sum_constr_equal}
 \end{align}
\end{itemize}

By changing $P^\star(3|\ve z^L)$ and $P^\star(5|\ve z^L)$ under the sum-constraint \eqref{eq:sum_constr} one can obtain the maximal and minimal link capacities for $c_{24}^{(j)\star}$ and $c_{34}^{(j)\star}$, as follows:
\begin{enumerate}
 \item $c_{34}^{(j)\star}=0$, $c_{24}^{(j)\star} = D_j - c_{14}^{(j)}$, $j=\{1,2\}$, by setting $P^\star(3|\ve z^L)=P(3|\ve z^L)+P(5|\ve z^L)$ and $P^\star(5|\ve z^L)=0$, $\forall \ve z^L \in \mc Z^L$.
 \item $c_{24}^{(j)\star}=0$, $c_{34}^{(j)\star} = D_j - c_{14}^{(j)}$, $j=\{1,2\}$, by setting $P^\star(3|\ve z^L)=0$ and $P^\star(5|\ve z^L)=P(3|\ve z^L)+P(5|\ve z^L)$,  $\forall \ve z^L \in \mc Z^L$.
\end{enumerate}
The link capacities are linear with respect to $P^\star_{A_t|\ve Z_{t-L}^{t-1}}$, hence any convex combination of $\left( D_j - c_{14}^{(j)}, 0 \right)$ and $\left(0, D_j - c_{14}^{(j)}\right)$ can be obtained for $\left( c_{24}^{(j)\star},c_{34}^{(j)\star}\right)$. 

We will now show that choosing $P^\star_{A_t|\ve Z_{t-L}^{t-1}}$ as in \eqref{eq:pstar1}~-~\eqref{eq:sum_constr} suffices to ensure the desired criteria \eqref{eq:cuts_criterion1}~-~\eqref{eq:cuts_criterion2}.
We have to distinguish two cases: 

 \subsubsection*{Case I} $A_j \leq D_j$ for at least one $j \in \{1,2\}$.\\
 This happens if we have
 \begin{align}
  c_{12}^{(j)} + c_{13} \leq c_{24}^{(j)} + c_{34}^{(j)} \quad \text{ for some } j \in \{1,2\}. \label{eq:case1crit}
 \end{align}
 In this case, we choose $P^\star(5|\ve z^L)$ such that $c_{34}^{(j)\star} = c_{13} - c_{32}^{(j)\star}$.
 By Lemma~\ref{lem:flow_div_equality}, this is sufficient to ensure the criteria \eqref{eq:cuts_criterion1}~-~\eqref{eq:cuts_criterion2}.\\
 We can always find such values for $P^\star(5|\ve z^L)$ because, by definition of Case I in \eqref{eq:case1crit},
 there is a $j\in\{1,2\}$ for which 
 \begin{align}
  D_j - c_{14}^{(j)} \geq c_{12}^{(j)} + c_{13}. \label{eq:case1_labeleq}
 \end{align}
 The LHS of \eqref{eq:case1_labeleq} is the maximal possible link capacity for $c_{34}^{(j)\star}$. The RHS of \eqref{eq:case1_labeleq} is larger than (or equal to) $c_{13} - c_{32}^{(j)\star}$, which is the desired value for $c_{34}^{(j)\star}$.
We can adjust $c_{34}^{(j)\star}$ between $0$ and $D_j - c_{14}^{(j)}$. As $c_{13} - c_{32}^{(j)\star}$ lies in this interval, we can choose values for $P^\star(5|\ve z^L)$ such that $c_{34}^{(j)\star} = c_{13} - c_{32}^{(j)\star}$. 
 \subsubsection*{Case II} $D_j \leq A_j$ for both $j =\{1,2\}$.\\
 In this case the sufficient criterion in Lemma~\ref{lem:flow_div_equality} cannot be guaranteed, since it is possible that  
there are no values for $P^\star(5|\ve z^L)$ such that $c_{13} = c_{32}^{(j)\star} + c_{34}^{(j)\star}$.\\
To satisfy the criteria \eqref{eq:cuts_criterion1}~-~\eqref{eq:cuts_criterion2}, we need for both $j =\{1,2\}$:
\begin{alignat}{2}
 D_j &\leq B_j^\star, \qquad& \text{hence} \qquad  c_{34}^{(j)\star} &\leq c_{13} \label{eq:CaseIIcrit1} \\
  D_j &\leq C_j^\star, & \text{hence} \qquad c_{24}^{(j)\star} &\leq c_{32}^{(j)\star} + c_{12}^{(j)} \label{eq:CaseIIcrit2}
\end{alignat}
In this case we choose $P^\star(5|\ve z^L)$ such that $\max \left\{c_{34}^{(1)\star}, c_{34}^{(2)\star} \right\}$ is as large as possible, but at most equal to $c_{13}$, in order not to violate \eqref{eq:CaseIIcrit1}. Two sub-cases need to be distinguished:
 \subsubsection*{Case IIa} One can choose $P^\star(5|\ve z^L)$ such that $\max \left\{c_{34}^{(1)\star}, c_{34}^{(2)\star} \right\} = c_{13}$.\\
 The condition in \eqref{eq:CaseIIcrit1} is satisfied by construction, so we have to check only \eqref{eq:CaseIIcrit2}.
 Note that the following inequalities always hold for Case IIa:
 \begin{align}
  c_{24}^{(j)} + c_{34}^{(j)} &\leq c_{12}^{(j)} + c_{13}, \qquad \text{because } D_j \leq A_j, \text{ and} \nonumber \\
  c_{24}^{(j)\star} + c_{34}^{(j)\star} &\leq c_{12}^{(j)} + c_{13}, \qquad \text{because of \eqref{eq:sum_constr_equal}. Hence, }\nonumber \\
  c_{24}^{(j)\star}  &\leq c_{12}^{(j)} + \left( c_{13} - c_{34}^{(j)\star} \right).
 \end{align}
 For condition \eqref{eq:CaseIIcrit2} to hold, the term $\left( c_{13} - c_{34}^{(j)\star} \right)$ should be smaller than (or equal to) $c_{32}^{(j)\star}$ for both $j=\{1,2\}$. \\
 For $j_{\max} = \argmax_{j \in \{1,2\}} c_{34}^{(j)\star} $, we have $ c_{13} - c_{34}^{(j_{\max})\star}  = 0$, satisfying the condition. \\
 For $j_{\min}= \argmin_{j \in \{1,2\}} c_{34}^{(j)\star} $, we have $c_{13} - c_{34}^{(j_{\min})\star}  \geq 0$. We next show that  $c_{13} - c_{34}^{(j_{\min})\star}  \leq c_{32}^{(j_{\min})\star}$ holds for this case as well.
 As $c_{13} = c_{34}^{(j_{\max})\star}$, we have
 \begin{align}
  &c_{13} - c_{34}^{(j_{\min})\star} = c_{34}^{(j_{\max})\star} - c_{34}^{(j_{\min})\star} \nonumber\\
  =&\sum_{\ve z^L \in \mc Z^L} P(\ve z^L) \left( \epsilon_{j_{\min}}(\ve z^L) - \epsilon_{j_{\max}}(\ve z^L)\right) P^\star(5|\ve z^L) \geq 0. \nonumber
 \end{align}
The following statement shows that $c_{13} - c_{34}^{(j_{\min})\star}  \leq c_{32}^{(j_{\min})\star}$:
\begin{align}
 &c_{13} - c_{34}^{(j_{\min})\star} - c_{32}^{(j_{\min})\star} \nonumber \\
 = & \sum_{\ve z^L \in \mc Z^L} P(\ve z^L) \left[ \epsonetwo{\ve z^L} - \epsilon_{j_{\max}}(\ve z^L)\right]P^\star(5|\ve z^L) \leq 0
\end{align}
as $\epsonetwo{\ve z^L} - \epsilon_{j_{\max}}(\ve z^L)\leq 0$ for all $\ve z^L \in \mc Z^L$, $j\in \{1,2\}$.
This shows that the choice of $P^\star(5|\ve z^L)$ such that $\max \left\{c_{34}^{(1)\star}, c_{34}^{(2)\star} \right\} = c_{13}$ is sufficient to achieve the criteria \eqref{eq:cuts_criterion1}~-~\eqref{eq:cuts_criterion2} in this case.
\subsubsection*{Case IIb} Choosing the maximal $P^\star(5|\ve z^L) = P(3|\ve z^L)+P(5|\ve z^L)$ $\forall \ve z^L \in \mc Z^L$ leads to $\max \left\{c_{34}^{(1)\star}, c_{34}^{(2)\star} \right\} < c_{13}$.\\
This immediately satisfies the condition in \eqref{eq:CaseIIcrit1}.
As $P^\star(5|\ve z^L)$ is maximal, we have $P^\star(3|\ve z^L)=0$ $\forall \ve z^L \in \mc Z^L$. Hence,\linebreak $c_{24}^{(j)\star}=0$, $\forall j \in \{1,2\}$, satisfying the condition \eqref{eq:CaseIIcrit2}.
This completes the proof.

\section{Proof of Proposition \ref{prop:maxweight_new_hmm}}
\label{ap-proof_prop_maxweight_hmm}
With slight abuse of notation, let $Q_{l,t}^{(j)}$ denote the number of packets stored in buffer $Q_{l}^{(j)}$ at time $t$. Obviously, $Q_{l,t}^{(j)} \in \mathbb N_0$. 
Define
\begin{align}
 \ve Q_t = \left( Q_{1,t}^{(1)}, Q_{2,t}^{(1)}, Q_{3,t}^{(1)}, Q_{1,t}^{(2)}, Q_{2,t}^{(2)},Q_{3,t}^{(2)} \right) \in \mathbb N_0^6.
\end{align}
Because $Q_4^{(1)}= Q_4^{(2)}=0$ by definition, the vector $\ve Q_t$ determines the queue state at time $t$.

Let $F_{rl,t}^{(j)}$ denote the number of packets that can travel from buffer $Q_r^{(j)}$ to $Q_l^{(j)}$ in time slot $t$.
Clearly, we have $F_{rl,t}^{(j)} \in \{0,1\}$. $F_{rl,t}^{(j)}$ is a deterministic function of the action $A_t$ and on the random erasure events $\ve Z_t$. Because of the erasures, the values of $F_{rl,t}^{(j)}$ are unknown to the transmitter before the transmission.
The \emph{actual} number of packets travelling on the link from buffer $Q_r^{(j)}$ to $Q_l^{(j)}$ in time slot $t$ is denoted $\tilde F_{rl,t}^{(j)}$. $\tilde F_{rl,t}^{(j)}$ may be different from $F_{rl,t}^{(j)}$, e.g. because buffer $Q_l^{(j)}$ is empty and no packet can be transmitted. 
$\tilde F_{rl,t}^{(j)}$ is a deterministic function of $A_t$, $\ve Z_t$ and $\ve Q_t$. 
We have $\tilde F_{rl,t}^{(j)} = \truth\{Q_{r,t}^{(j)}>0\} F_{rl,t}^{(j)}$,
hence
$
 \tilde F_{rl,t}^{(j)} \leq F_{rl,t}^{(j)}.
$

To model external packets arrivals,
let $F_{01,t}^{(j)}= \tilde F_{01,t}^{(j)}$ denote the indicator random variable if a packet arrived in queue $Q_1^{(j)}$ during time slot $t$. $F_{01,t}^{(j)}$ is independent of all other random variables in the system, and we have
\begin{align}
\mathbb E[F_{01,t}^{(j)}]=R_j .
\end{align}

The \emph{flow divergence} \cite[Chapter 1.1.2]{bertsekas1998network} 
at buffer $Q_l^{(j)}$ is defined as
\begin{align}
\flowdivRV{l,t}{(j)} = \sum_{m} F_{lm,t}^{(j)} - \sum_{k} F_{kl,t}^{(j)}. \label{eq:def_flow_div}
\end{align}
The flow divergence is thus the difference of the number of packets that can depart from buffer $Q_l^{(j)}$ minus the number of packets that can arrive at $Q_l^{(j)}$.

The number of packets in queue $Q_l^{(j)}$ evolve according to 
\begin{align}
 Q_{l,t+1}^{(j)} &= Q_{l,t}^{(j)} - \sum_{m} \tilde F_{rl,t}^{(j)}  +\sum_{k} \tilde F_{kl,t}^{(j)} \nonumber\\
&\leq \left[Q_{l,t}^{(j)} - \sum_{m} F_{rl,t}^{(j)} \right]^+ +\sum_{k} F_{kl,t}^{(j)}, \label{eq:queue_dynamics}
\end{align}
where $[x]^+ = \max(x,0)$.
There is an inequality rather than an equality because some queue $Q_k^{(j)}$ might be empty, hence $\tilde F_{kl,t}^{(j)} < F_{kl,t}^{(j)}$.

Each coding scheme that we will define leads to a different evolution of the queue state $\ve Q_t$, as the actions at time $t$ might be different.
Our main interest is whether all queues in the network are stable.
A queue $Q_t$ is strongly stable if \cite{neely2010stochastic}
\begin{align}
 \limsup_{n\rightarrow \infty} \frac{1}{n} \sum_{t=1}^n \mathbb{E}[Q_t] < \infty. \label{eq:strong_stability}
\end{align}
A network of queues is strongly stable if all queues inside the network are strongly stable.

\begin{figure}[t]
\centering
  \tikzset{>=latex}
\begin{tikzpicture}[scale = 0.7, every node/.style={scale=0.7}]
\input{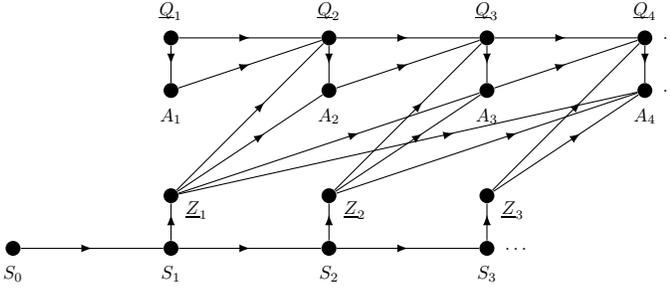}
 \end{tikzpicture}
\caption{Bayesian network of the queuing system. 
Actions at time $t$ are permitted to depend on the previous feedback messages $\ve Z^{t\hspace{-.025cm}-\hspace{-.025cm}1}$ and the current buffer state $\ve Q_t$. 
 Note that $\ve Q_t$ denotes the buffer state \emph{before} action $A_t$ is executed.  }
\label{fig:fdg_queues_hidden}
\end{figure}

The action at time $t$ can depend only on the current queue state $\ve Q_t$ and on the previous channel feedback messages $\ve Z^{t\hspace{-.025cm}-\hspace{-.025cm}1}$. Hence it can be represented by a distribution $P_{A_t|\ve Q_t \ve Z^{t\hspace{-.025cm}-\hspace{-.025cm}1}}$.
All dependencies are depicted in the Bayesian network in Fig.~\ref{fig:fdg_queues_hidden}.
Define the Lyapunov function $\mathbb L(\ve Q_t)$ as
\begin{align}
 \mathbb L(\ve Q_t) = \sum_{j=1}^2 \sum_{l=1}^3 \left( Q_{l,t}^{(j)} \right)^2
\end{align}
and the $T$-slot conditional Lyapunov drift $\Delta(\ve Q_t)$ as
\begin{align}
 \Delta(\ve Q_t) = \mathbb{E}\left[ \mathbb L(\ve Q_{t+T}) - \mathbb L(\ve Q_{t}) \left|\ve Q_t \right. \right]. \label{eq:condLyap_hmm}
\end{align}
The expectation is with respect to the possibly random actions $A_t,A_{t+1},\ldots,A_{t+T\hspace{-.025cm}-\hspace{-.025cm}1}$, erasures $\ve Z_t, \ve Z_{t+1},\ldots,\ve Z_{t+T\hspace{-.025cm}-\hspace{-.025cm}1}$ and queues $\ve Q_{t+1},\ve Q_{t+2},\ldots,\ve Q_{t+T}$.
$\Delta(\ve Q_t)$ measures the expected reduction or increase of the queue lengths from slot $t$ to slot $t+T$, for some fixed constant $T$, conditioned on $\ve Q_t$, and is useful to prove strong stability.

We split the $T$-slot conditional Lyapunov drift $\Delta(\ve Q_t)$  into the telescoping sum
\begin{align}
  \Delta(\ve Q_t) \!=\!& \sum_{\tau=t}^{t+T\hspace{-.025cm}-\hspace{-.025cm}1} \mathbb{E}\left[ \mathbb L(\ve Q_{\tau+1}) - \mathbb L(\ve Q_\tau) \left| \ve Q_t \right. \right]
  \label{eq:telescopingsum_hmm}\\
\!=\!& \sum_{\tau=t}^{t+T\hspace{-.025cm}-\hspace{-.025cm}1} \mathbb E \Big[ \mathbb E \left[ \mathbb L(\ve Q_{\tau+1}) - \mathbb L(\ve Q_\tau) \big| \ve Q_\tau \ve Z^{\tau\hspace{-.025cm}-\hspace{-.025cm}1}  \right] \Big| \ve Q_t \Big], \label{eq:total_exp_hmm}
\end{align}
where the last equality is because $\ve Q_{\tau+1}$ is a deterministic function of $\ve Q_{\tau}, A_\tau, \ve Z_\tau$ and $ \ve Q_{\tau} A_\tau \ve Z_\tau - \ve Q_\tau \ve Z_{\tau-L}^{\tau\hspace{-.025cm}-\hspace{-.025cm}1} - \ve Q_t$ forms a Markov chain for $\tau\geq t$.

We bound the individual terms inside the inner expectation of \eqref{eq:total_exp_hmm} by using \cite[Lemma 4.3]{georgiadis2006resource}, which states that
for any nonnegative numbers $v,u,\mu,\alpha$ satisfying $v \leq [u-\mu]^+ + \alpha$, we have
\begin{align}
v^2 &\leq u^2 + \mu^2 + \alpha^2 - 2u(\mu - \alpha).
\end{align}
We apply this lemma and combine it with $(F_{lm,\tau}^{(j)})^2 = F_{lm,\tau}^{(j)}$ because $F_{lm,\tau}^{(j)}$ is either $1$ or $0$ and obtain: 
 \begin{align}
 \mathbb L(\ve Q_{\tau+1})-\mathbb L(\ve Q_\tau)
 &\leq 
 12 - 2\sum_{j=1}^2   \sum_{l=1}^3 Q_{l,\tau}^{(j)} \flowdivRV{l,\tau}{(j)}
 \label{eq:flow_bound}
\end{align}
where $\flowdivRV{l,\tau}{(j)}$ is the flow divergence defined in \eqref{eq:def_flow_div}.

We use \eqref{eq:total_exp_hmm} together with \eqref{eq:flow_bound} to obtain 
 \begin{align}
 \Delta(\ve Q_t)  &\!\stackrel{}{\leq}\!\!\! \sum_{\tau=t}^{t+T\hspace{-.025cm}-\hspace{-.025cm}1}\!\!\mathbb E\!\!\left[\mathbb E\!\!\left[\!12\! -\! 2\!\sum_{j=1}^2 \!  \sum_{l=1}^3 \!Q_{l,\tau}^{(j)} \flowdivRV{l,\tau}{(j)} \bigg| \ve Q_\tau \ve Z^{\tau\hspace{-.025cm}-\hspace{-.025cm}1} \!\hspace{-.05cm} \right] \!\!\bigg| \ve Q_t\! \right]\!\!. \label{eq:drift_eq_hmm}
\end{align}
The terms in the inner expectation are
\begin{align}
 \mathbb E\!\left[\!\flowdivRV{1,\tau}{(j)}\big| \ve Q_\tau \ve Z^{\tau\hspace{-.025cm}-\hspace{-.025cm}1}\!\right] &\!\!=\!\! \left(\!\!\!\begin{array}{l}\Pr[A_t\!\!=\!j|\ve Q_\tau \ve Z^{\tau\hspace{-.025cm}-\hspace{-.025cm}1}] \\\!+\! \Pr[A_t\!\!=\!4|\ve Q_\tau \ve Z^{\tau\hspace{-.025cm}-\hspace{-.025cm}1}] \!\!\!\hspace{-.05cm}\end{array}\right) \!\!\big(\hspace{-.05cm}1\!\!-\!\epsonetwo{\ve Z^{\tau\hspace{-.025cm}-\hspace{-.025cm}1}}\!\big) \!\!-\! \!R_j \label{eq:exp_flow_div_Q1_hmm}\\
 \mathbb E\!\left[\!\flowdivRV{2,\tau}{(j)}\big| \ve Q_\tau \ve Z^{\tau\hspace{-.025cm}-\hspace{-.025cm}1}\!\right] &\!\!=\!\!\Pr[A_t\!=\!3|\ve Q_\tau \ve Z^{\tau\hspace{-.025cm}-\hspace{-.025cm}1}] \big(1-\epsj{\ve Z^{\tau\hspace{-.025cm}-\hspace{-.025cm}1}}\big) \nonumber\\
 -&\!\! \left(\!\!\!\begin{array}{l}\Pr[A_t\!=\!j|\ve Q_\tau \ve Z^{\tau\hspace{-.025cm}-\hspace{-.025cm}1}]  \\\!+\! \Pr[A_t\!=\!5|\ve Q_\tau \ve Z^{\tau\hspace{-.025cm}-\hspace{-.025cm}1}\hspace{-.05cm}] \!\!\!\hspace{-.05cm}\end{array}\right)\!\!\big(\!\epsj{\hspace{-.05cm}\ve Z^{\tau\hspace{-.025cm}-\hspace{-.025cm}1}\hspace{-.05cm}} \!\!-\!\hspace{-.05cm} \epsonetwo{\hspace{-.05cm}\ve Z^{\tau\hspace{-.025cm}-\hspace{-.025cm}1}\hspace{-.05cm}}\!\big) \label{eq:exp_flow_div_Q2_hmm}\\
 \mathbb E\!\left[\!\flowdivRV{3,\tau}{(j)}\big| \ve Q_\tau \ve Z^{\tau\hspace{-.025cm}-\hspace{-.025cm}1}\!\right] &\!\!=\!\! \left(\!\!\!\begin{array}{l}\Pr[A_t\!\!=\!5|\ve Q_\tau \ve Z^{\tau\hspace{-.025cm}-\hspace{-.025cm}1}] \\\!-\! \Pr[A_t\!\!=\!4|\ve Q_\tau \ve Z^{\tau\hspace{-.025cm}-\hspace{-.025cm}1}] \!\!\!\hspace{-.05cm}\end{array}\right) \!\!\big(\hspace{-.05cm}1\!\!-\!\epsonetwo{\ve Z^{\tau\hspace{-.025cm}-\hspace{-.025cm}1}}\!\big), \label{eq:exp_flow_div_Q3_hmm}
\end{align}
where the expectation is with respect to a distribution $P_{A_\tau|\ve Q_\tau \ve Z^{\tau\hspace{-.025cm}-\hspace{-.025cm}1}}$.
All relations follow because $\ve Z_\tau - \ve Z^{\tau\hspace{-.025cm}-\hspace{-.025cm}1} - A_\tau \ve Q_\tau$ forms a Markov chain.

The strategy in Table~\ref{tab:new_det_algo_hmm} finds the tightest upper bound in \eqref{eq:drift_eq_hmm} under the assumption that actions can depend on $\ve Q_\tau$ and $\ve Z^{\tau\hspace{-.025cm}-\hspace{-.025cm}1}$. Hence, any stationary probabilistic scheme that bases its decisions only on $\ve Z_{\tau-L}^{\tau\hspace{-.025cm}-\hspace{-.025cm}1}$, according to a distribution $P_{A_\tau|\ve Z_{\tau-L}^{\tau\hspace{-.025cm}-\hspace{-.025cm}1}}$, leads to a looser upper bound on $\Delta(\ve Q_t)$.

We introduce notation for the average flow divergence:
\begin{align}
 \mathbb E\left[\flowdivRV{l,\tau}{(j)}\big| \ve Q_\tau = \ve q, \ve Z^{\tau\hspace{-.025cm}-\hspace{-.025cm}1} = \ve z^{\tau\hspace{-.025cm}-\hspace{-.025cm}1}\right] = \flowdiv{l}{(j)}{(\ve z_{\tau-L}^{\tau\hspace{-.025cm}-\hspace{-.025cm}1})}
\end{align}
where the expectation on the LHS is with respect to a distribution $P_{A_\tau|\ve Z_{\tau-L}^{\tau\hspace{-.025cm}-\hspace{-.025cm}1}}$. This expression does not depend on the queue state $\ve Q_\tau$ and $\ve Z^{\tau-L-1}$.

The relations in \eqref{eq:exp_flow_div_Q1_hmm} - \eqref{eq:exp_flow_div_Q3_hmm} are adapted to
\begin{align*}
 \flowdiv{1}{(j)}{(\ve Z_{\tau-L}^{\tau\hspace{-.025cm}-\hspace{-.025cm}1})} &= \left(\!\!\!\begin{array}{l}\Pr[A_t\!=\!j|\ve Z_{\tau-L}^{\tau\hspace{-.025cm}-\hspace{-.025cm}1}] \\+ \Pr[A_t\!=\!4|\ve Z_{\tau-L}^{\tau\hspace{-.025cm}-\hspace{-.025cm}1}]\end{array} \!\!\!\right) \!\big(1-\epsonetwo{\ve Z_{\tau-L}^{\tau\hspace{-.025cm}-\hspace{-.025cm}1}}\big) \!-\! R_j \\
\flowdiv{2}{(j)}{(\ve Z_{\tau-L}^{\tau\hspace{-.025cm}-\hspace{-.025cm}1})} &= 
 \Pr[A_t\!=\!3|\ve Z_{\tau-L}^{\tau\hspace{-.025cm}-\hspace{-.025cm}1}] \big(1-\epsj{\ve Z_{\tau-L}^{\tau\hspace{-.025cm}-\hspace{-.025cm}1}}\big) \nonumber\\
 -& \left(\!\!\!\begin{array}{l}\Pr[A_t\!=\!j|\ve Z_{\tau-L}^{\tau\hspace{-.025cm}-\hspace{-.025cm}1}] \\ + \Pr[A_t\!=\!5|\ve Z_{\tau-L}^{\tau\hspace{-.025cm}-\hspace{-.025cm}1}] \end{array}\!\!\!\right)\!\big(\epsj{\ve Z_{\tau-L}^{\tau\hspace{-.025cm}-\hspace{-.025cm}1}} - \epsonetwo{\ve Z_{\tau-L}^{\tau\hspace{-.025cm}-\hspace{-.025cm}1}}\big) \\
\flowdiv{3}{(j)}{(\ve Z_{\tau-L}^{\tau\hspace{-.025cm}-\hspace{-.025cm}1})} &= \left(\!\!\!\begin{array}{l}\Pr[A_t\!=\!5|\ve Z_{\tau-L}^{\tau\hspace{-.025cm}-\hspace{-.025cm}1}]\\ - \Pr[A_t\!=\!4|\ve Z_{\tau-L}^{\tau\hspace{-.025cm}-\hspace{-.025cm}1}] \end{array}\!\!\!\right) \big(1-\epsonetwo{\ve Z_{\tau-L}^{\tau\hspace{-.025cm}-\hspace{-.025cm}1}}\big).
\end{align*}

Hence we can continue with the chain of inequalities in \eqref{eq:drift_eq_simpl_hmm}~-~\eqref{eq:final_bound_Delta_hmm} on the top of the page.
\begin{figure*}
 \begin{align}
 \Delta(\ve Q_t) 
 &\leq \sum_{\tau=t}^{t+T\hspace{-.025cm}-\hspace{-.025cm}1} 12 - \mathbb E \left[ 2\sum_{j=1}^2  \sum_{l=1}^3 Q_{l,\tau}^{(j)} \flowdiv{l}{(j)}{(\ve Z_{\tau-L}^{\tau\hspace{-.025cm}-\hspace{-.025cm}1})}\bigg|\ve Q_t \right] \label{eq:drift_eq_simpl_hmm}\\
 &\stackrel{(a)}{\leq} \sum_{\tau=t}^{t+T\hspace{-.025cm}-\hspace{-.025cm}1} 12 + 12 (\tau-t) - \mathbb E \left[ 2\sum_{j=1}^2 \sum_{l=1}^3  Q_{l,t}^{(j)} \flowdiv{l}{(j)}{(\ve Z_{\tau-L}^{\tau\hspace{-.025cm}-\hspace{-.025cm}1})} \bigg|\ve Q_t \right] \label{eq:drift_eq_with_Qbound_hmm}\\
 &\stackrel{(b)}{\leq} 12T + 6T^2 - 2 \sum_{j=1}^2 \sum_{\tau=t}^{t+T\hspace{-.025cm}-\hspace{-.025cm}1} \sum_{\ve z_{\tau-L}^{\tau\hspace{-.025cm}-\hspace{-.025cm}1}} \Pr\left[\ve Z_{\tau-L}^{\tau\hspace{-.025cm}-\hspace{-.025cm}1}=\ve z_{\tau-L}^{\tau\hspace{-.025cm}-\hspace{-.025cm}1}|\ve Q_t\right]  \sum_{l=1}^3   Q_{l,t}^{(j)} \flowdiv{l}{(j)}{(\ve z_{\tau-L}^{\tau\hspace{-.025cm}-\hspace{-.025cm}1})} \nonumber \\
 &\stackrel{(c)}{=}12T + 6T^2- 2 \sum_{j=1}^2     \sum_{l=1}^3  Q_{l,t}^{(j)} \sum_{\ve z^L}  \flowdiv{l}{(j)}{(\ve z^L)}  \sum_{\tau=t}^{t+T\hspace{-.025cm}-\hspace{-.025cm}1} \Pr[\ve Z_{\tau-L}^{\tau\hspace{-.025cm}-\hspace{-.025cm}1}=\ve z^L|\ve Q_t]\\
 &\stackrel{(d)}{\leq} 12T+6T^2 - 2T\sum_{j=1}^2 \sum_{l=1}^3 Q_{l,t}^{(j)} \left( \left[\sum_{\ve z^L}  P_{\ve Z^{L}}(\ve z^L) \flowdiv{l}{(j)}{(\ve z^L)} \right] - \varepsilon_{L,T}    \right) \\
 &\stackrel{(e)}{\leq}12T + 6T^2 -2T(\delta-\varepsilon_{L,T}) \sum_{j=1}^2 \sum_{l=1}^3 Q_{l,t}^{(j)} .\label{eq:final_bound_Delta_hmm}
\end{align}
\vspace*{-4mm}
\end{figure*}

For step $(a)$ we have used the property that buffer level $Q_{l,\tau}^{(j)}$ can decrease by at most one per slot.
The expression inside the expectation in \eqref{eq:drift_eq_with_Qbound_hmm} does not depend on $\ve Q_\tau$ anymore.
Step $(b)$ writes out the expectation and rearranges terms.
Because the sequence $\ve Z^n$ is stationary and also the probabilistic strategy is stationary, $\flowdiv{l}{(j)}{(\ve z_{\tau-L}^{\tau\hspace{-.025cm}-\hspace{-.025cm}1})}$ does not depend on $\tau$ but only on the realization $\ve Z_{\tau-L}^{\tau\hspace{-.025cm}-\hspace{-.025cm}1} = \ve z^L$. This is used in step $(c)$.

Define the mixture distribution 
\begin{align}
 M_T(\ve z^L|\ve q) = \frac{1}{T} \sum_{\tau=t}^{t+T\hspace{-.025cm}-\hspace{-.025cm}1} P_{\ve Z_{\tau-L}^{\tau\hspace{-.025cm}-\hspace{-.025cm}1}|\ve Q_t}(\ve z^L|\ve q ).
\end{align}

The constant $T$ can be chosen large enough such that
 $ \left| \sum_{\ve z^L} \flowdiv{l}{(j)}{(\ve z^L)} \left( P_{\ve Z^{L}}(\ve z^L) - M_T(\ve z^L|\ve q) \right)  \right|$
is small, for all $l$, $j$ and $\ve q$.
This follows because
\begin{align}
&\left| \sum_{\ve z^L} \flowdiv{l}{(j)}{(\ve z^L)} \left( P_{\ve Z^L}(\ve z^L) - M_T(\ve z^L|\ve q) \right)  \right| \nonumber \\
\leq & \sum_{\ve z^L} \left| \flowdiv{l}{(j)}{(\ve z^L)} \right|  \left| P_{\ve Z^L}(\ve z^L) - M_T(\ve z^L|\ve q)  \right| \nonumber \\
\leq & \sum_{\ve z^L}   \left| P_{\ve Z^L}(\ve z^L) - M_T(\ve z^L|\ve q)  \right| 
\leq \varepsilon_{L,T}, \quad \forall\ve q \label{eq:converge_to_steady_hmm}
\end{align}
for some $\varepsilon_{L,T} >0$. Hence, the variational distance between $P_{\ve Z^L}$ and $M_T$ needs to become small.

Recall that the channel state Markov chain is irreducible and aperiodic. 
Define the random variable $U_t = (S_{t-L}^{t\hspace{-.025cm}-\hspace{-.025cm}1} \ve Z_{t-L}^{t\hspace{-.025cm}-\hspace{-.025cm}1})$. Note that $U_t$ is Markovian and the corresponding Markov chain is irreducible and aperiodic.
One may verify that $U_{\tau+1}$ is independent of $\ve Q_t$ given $U_{\tau}$ for $\tau \geq t$ and hence the distribution of $U_\tau$ converges to the steady state distribution of the corresponding Markov chain, for any initial distribution of $P_{U_t|\ve Q_t}$. 
If $U_\tau$ is distributed according to the steady state distribution, so is its component $\ve Z_{\tau-L}^{\tau\hspace{-.025cm}-\hspace{-.025cm}1}$, i.e. according to $P_{\ve Z_{\tau-L}^{\tau\hspace{-.025cm}-\hspace{-.025cm}1}}(\ve z^L)$. This distribution does not depend on $\tau$ because $\ve Z^n$ is stationary, so we can replace it with $P_{\ve Z^L}(\ve z^L)$.
If $P_{\ve Z_{\tau-L}^{\tau\hspace{-.025cm}-\hspace{-.025cm}1}|\ve Q_t}$ converges to $P_{\ve Z^L}$, so does the Ces\`{a}ro mean in \eqref{eq:converge_to_steady_hmm}. In general the constant $T$ must be significantly larger than $L$.

For step $(d)$, we replace the expression $\sum_{\ve z^L} \flowdiv{l}{(j)}{(\ve z^L)} \sum_{\tau=t}^{t+T\hspace{-.025cm}-\hspace{-.025cm}1} \Pr\left[\ve Z_{\tau-L}^{\tau\hspace{-.025cm}-\hspace{-.025cm}1}=\ve z^L|\ve Q_t\right]$ by its lower bound $T\left(\sum_{\ve z^L} \flowdiv{l}{(j)}{(\ve z^L)} P_{\ve Z^L}(\ve z^L)\right)-T\varepsilon_{L,T}$.

For step $(e)$, note that if the rate pair is in the interior of the $L^{\text{th}}$ order approximation of the outer bound, i.e. if \linebreak 
$(R_1+\bar{\delta}, R_2+\bar{\delta}) \in \bar{\mc C}_{\text{ fb}}^\text{mem}(L)$, then there exists a constant $\delta >0$ that goes to zero when $\bar{\delta} \rightarrow 0$ such that
\begin{align}
\sum_{\ve z^L}  P_{\ve Z^L}(\ve z^L) \flowdiv{l}{(j)}{(\ve z^L)}\geq \delta, \quad \forall~l\in \{1,2,3\},~j\in \{1,2\}, \nonumber
\end{align}
where $\delta$ should be chosen such that $\delta > \varepsilon_{L,T}$.

Using the result in \eqref{eq:final_bound_Delta_hmm} and the law of total expectation, we can bound
\begin{align}
 \mathbb E \left[ \mathbb L(\ve Q_{t+T}) - \mathbb L(\ve Q_t) \right] = &\mathbb E \big[ \Delta(\ve Q_t) \big| \ve Q_t \big] \nonumber \\
  \leq  18T^2 - 2T(\delta-\varepsilon_{L,T} ) & \sum_{j=1}^2 \sum_{l=1}^3 \mathbb E\left[ Q_{l,t}^{(j)}  \right].
\end{align}
Summing over all time slots $t=1,\ldots,n$ yields
\begin{align}
\sum_{t=1}^{n} &\mathbb E\left[ \mathbb L(\ve Q_{t+T})- \mathbb L(\ve Q_t) \right] = \sum_{t=1}^T \mathbb E\left[\mathbb L(\ve Q_{t+n}) - \mathbb L(\ve Q_t)  \right]  \nonumber \\\leq &18 n T^2 - 2 T (\delta-\varepsilon_{L,T} )   \sum_{t=1}^{n} \sum_{j=1}^2 \sum_{l=1}^3 \mathbb E\left[Q_{l,t}^{(j)}  \right].
 \end{align}
 Rearranging terms gives
 \begin{align}
 \frac{1}{n} \sum_{t=1}^n \sum_{j=1}^2 \sum_{l=1}^3 \mathbb E[Q_{l,t}^{(j)} ] &\leq \frac{9T}{\delta - \varepsilon_{L,T} } + \frac{\sum_{t=1}^T \mathbb E[\mathbb L(\ve Q_{t}) ]}{2(\delta - \varepsilon_{L,T})T n},
\end{align}
and taking a $\limsup$ with respect to $n$ on both sides proves strong stability of the queuing network, given that $1/T\sum_{t=1}^T \mathbb E[\mathbb L(\ve Q_{t}) ]<\infty$. This is true for any finite constant $T$ and $\mathbb E[\mathbb L(\ve Q_1)]<\infty$. 
\vspace*{15mm}

\vspace*{-10mm}

\IEEEtriggeratref{14}
\bibliographystyle{IEEEtran}
\bibliography{biblio}

\end{document}